# One-Dimensional van der Waals Quantum Materials – State of the Art and Perspectives

**Alexander A. Balandin[1]♦, Fariborz Kargar[1], Tina T. Salguero[2], and Roger K. Lake[1]**

[1]Department of Electrical and Computer Engineering, Materials Science and Engineering Program, University of California, Riverside, California 92521 U.S.A.

[2]Department of Chemistry, University of Georgia, Athens, Georgia 30602 USA

## Abstract

The advent of graphene and other two-dimensional van der Waals materials, with their unique electrical, optical, and thermal properties, has resulted in tremendous progress for fundamental science. Recent developments suggest that taking one more step down in dimensionality — from monolayer, atomic sheets to individual atomic chains — can bring exciting prospects as the ultimate limit in material downscaling is reached while establishing an entirely new field of one-dimensional quantum materials. Here we review this emerging area of one-dimensional van der Waals quantum materials and anticipate its future directions. We focus on quantum effects associated with the charge-density-wave condensate, strongly-correlated phenomena, topological phases, and other unique physical characteristics, which are attainable specifically in van der Waals materials of lower dimensionality. Possibilities for engineering the properties of quasi-one-dimensional materials via compositional changes, vacancies, and defects, as well as the prospects of their applications in composites are also discussed.

♦ Corresponding author (AAB): balandin@ece.ucr.edu ; https://balandingroup.ucr.edu/

**Preface:** As they often say, science and history move in a spiral. In the context of nanotechnology, it is easy to identify the following important sequence (see Figure 1). Development of molecular beam epitaxy (MBE) brought the first electronic two-dimensional (2D) structures – quantum wells – with sufficiently small thicknesses and sharp interfaces to observe quantum coherent, quantum confined electrons, with all the exciting implications – from the quantum Hall effect to the high-electron-mobility transistors [1–3]. The vapor-liquid-solid (VLS) method and its modifications made available the one-dimensional (1D) structures – quantum wires – to make the transition to structures where the spatial confinement attains a new level [4,5]. Further progress in MBE, *i.e.* refinement of the Stranski–Krastanov growth [6,7], and the rise of colloidal chemistry [8–11] led to quantum dots, systems of zero-dimensionality (0D), that were hailed as "artificial atoms." The latter closed a cycle in the evolution of low-dimensional materials based on traditional or conventional crystalline semiconductors, primarily the III–V and II-VI materials. It is important to note that the term "quasi" should be associated with these 2D, 1D, and 0D materials because the semiconductor quantum wells, wires, and dots with thicknesses of hundreds of atomic planes, as well as strong covalent bonds between atomic planes, retain many features of the corresponding three-dimensional (3D) bulk crystals. Nevertheless, the cross-sectional dimensions of the MBE- or VLS- grown structures were just small enough, in comparison with the de Broglie wavelength of an electron, the electron mean free path (MFP), or the diameter of an exciton, to give the as-prepared samples some quantum flavor.

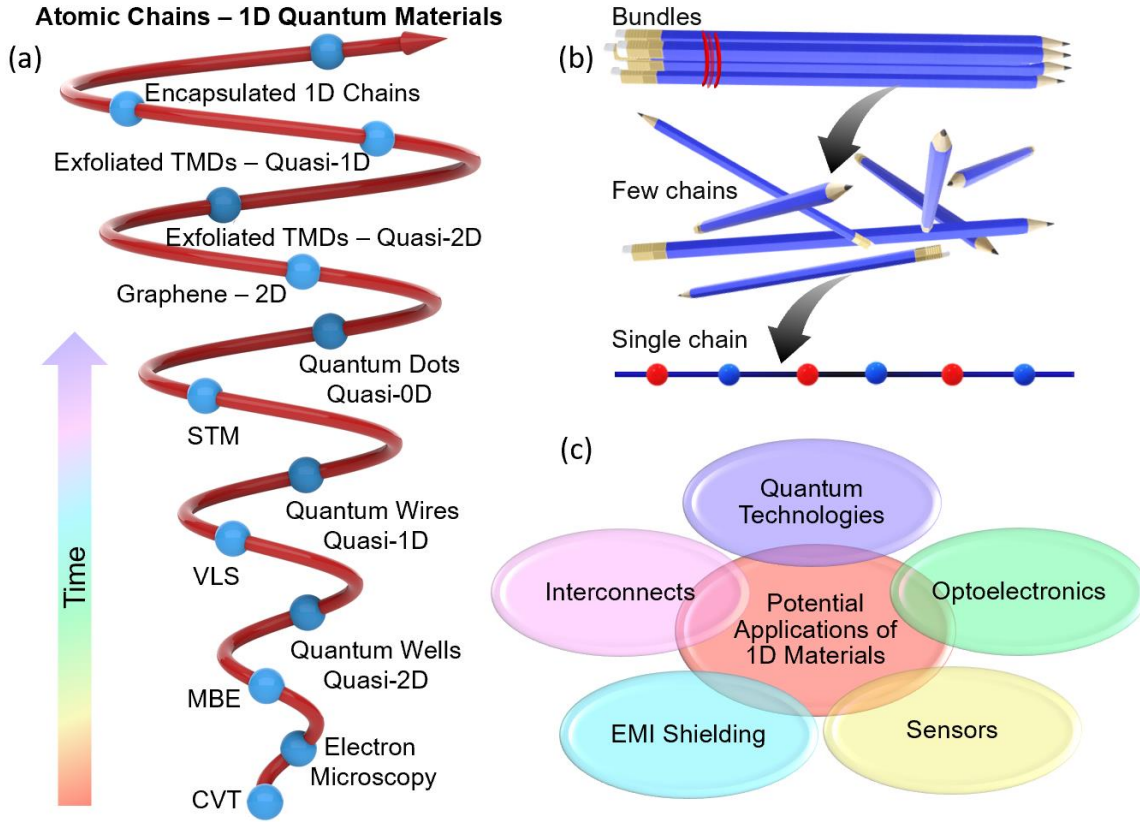

[**Figure 1:** Developments in the low-dimensional quantum materials. (a) Schematic showing the evolution of the low-dimensional materials driven by interest in emerging physical phenomena characteristic of quantum-confined systems and developments in synthesis and characterization techniques. (b) An illustration of exfoliation of the material with quasi-1D crystal structure to individual atomic chains. (c) Prospective applications of quasi-1D quantum materials.]

An overlapping but reverse cycle of low dimensional materials based specifically on carbon began with the discovery in 1985 of $C_{60}$, an approximately spherical shell of sixty carbon atoms, also referred to as a buckyball or a buckminsterfullerene due to its structural resemblance to the geodesic domes designed by Buckminster Fuller [12]. The $C_{60}$ clusters were produced by an arc-evaporation process. In 1991, elongated clusters, now known as single-wall carbon nanotubes (SWCNTs) were discovered among the products of the arc-evaporation [13,14]. The SWCNTs are 1D hollow tubes of C atoms with diameters down to 0.7 nm. One should note here that multi-wall carbon nanotubes (MWCNTs) were observed much earlier, in 1952 [15,16]. The SWCNTs were the subject of intense study until the mechanical exfoliation of graphene from graphite was reported in 2004 [17]. This event marks the start of the era of 2D van der Waals (vdW) materials, during which time the developments in low-dimensional carbon materials proceeded from 0D to 1D and, finally, to 2D materials. Graphite – a three-dimensional (3D) material that consists of 2D atomic planes weakly bonded by vdW forces – is a true bulk material that has been known for a long time, and it is not considered here.

Graphene, a monolayer sheet of carbon atoms, was initially denied existence based on theoretical thermodynamic considerations and misinterpretations of the Mermin-Wagner theorem (more about it below). In practice, 2D graphene was robustly exfoliated from 3D bulk graphite. Graphene is as close as one can get to the *true* 2D system [17–19]. Topologically, its only vestiges of the 3D world are the out-of-plane vibrational modes, *i.e*. ZA phonons, and wrinkles – manifestations of the three degrees of freedom always present in real structures [20–22]. The impact of graphene on physics and engineering has been extraordinary, proving that the dimensionality reduction can bring unprecedented material properties, including the extraordinary speed and mobility of Dirac electrons and size-divergent thermal conductivity of nearly-2D phonons.

In this Review, we describe the recent rapid emergence of a new research field based on 1D quantum materials. These are vdW materials with the 1D motifs in their crystal structure that are thinned down to, or grown as, individual *atomic chains* or few-chain *atomic threads* or *bundles*. These special cases come as close to 1D material systems as one can practically isolate and handle at ambient conditions. They are as different from conventional quantum wires as graphene is from MBE-grown quantum wells. The emergence of this latest generation of 1D quantum materials constitutes the next step forward in the continuous spiral of 3D –2D – 1D systems evolution (see Figure 1).

**Quasi-1D** *vs.* **True-1D Crystals:** The success of graphene has stimulated interest in other materials that can be mechanically or chemically exfoliated to obtain monolayer sheets. A large number of 2D materials in the family of transition metal dichalcogenides (TMDs) with the formula $MX_2$ (where M is a transition metal and X is S, Se, or Te), such as $MoS_2$, feature weak vdW bonding between planar structural units, allowing for exfoliation into quasi-2D layers [23–25].

The term "quasi" emphasizes that the structural unit of TMDs is not a single atomic plane like in graphene but an atomic tri-layer, *i.e.* S–Mo–S. In some cases, it also has been used to indicate that the structure has a thickness of more than one tri-layer and, in such a context, "quasi-2D" suggests a thin film of vdW material with the 2D layered crystal structure. One practical criterion for how thick the few-layer graphene or $MoS_2$ film can be while still being called a "quasi-2D" system is based on whether the Fermi level can be moved with a back gate, *i.e.* the carrier concentration can be changed.

There are numerous examples of TMDs that exfoliate into quasi-2D vdW layers in the scientific literature. The vdW materials based on 1D rather than 2D structural motifs are less well studied. However, one starts noticing a growing wave of interest in quasi-1D vdW materials research starting from about 2015 [26–32] [33–52]. Selected examples of crystal structures of materials from this class are illustrated in Figure 2. They include the family of transition metal trichalcogenides (TMTs) with the formula $MX_3$, *e.g.* $TaSe_3$, along with other metal chalcogenide compositions like $Nb_2Se_9$, metal halides like $NbCl_4$, and metal chalcohalides like $(TaSe_4)_2I$ [53–62].

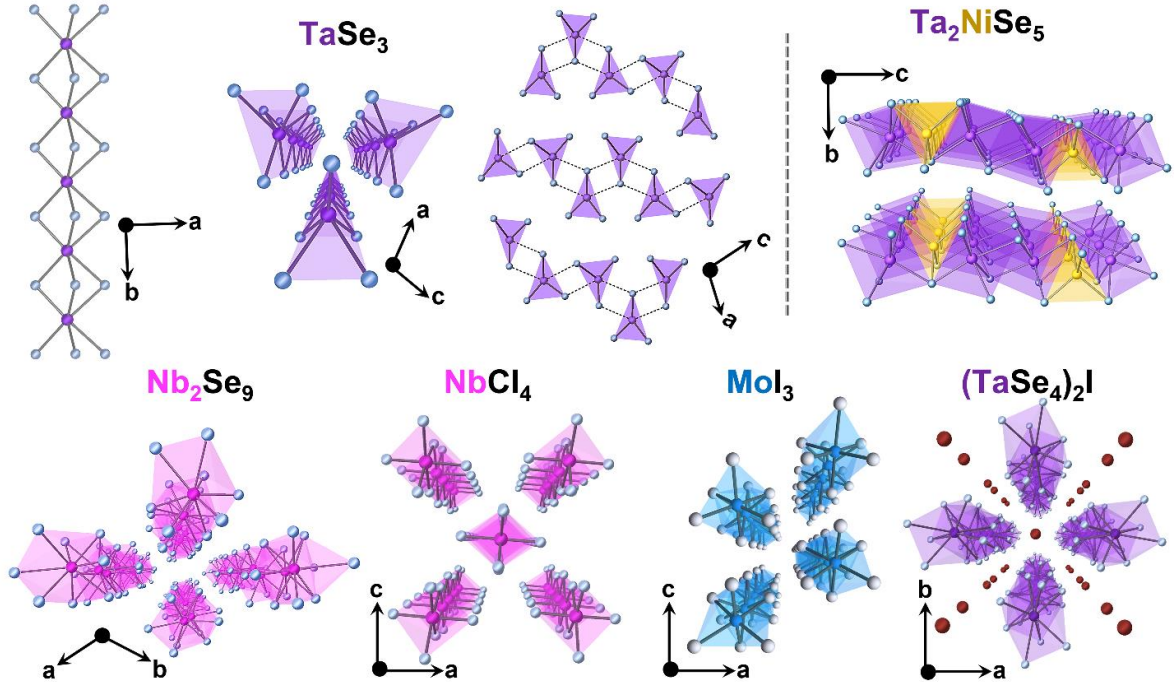

[**Figure 2:** Examples of quasi-1D and true-1D van der Waals crystal structures. The top panel shows quasi-1D vdW materials. These materials have strong covalent bonds in one direction while bonded in the perpendicular plane partially by weak vdW forces and partially by weaker covalent bonds than those along the atomic chains. The weaker covalent bonds are shown by dashed lines. The bottom panel shows true-1D vdW materials. These materials have strong covalent bonds in the direction of the atomic chain only; the rest of the bonds are of the weak vdW type.]

The TMTs have quasi-1D crystalline structures that lead to quasi-1D features in the electronic and phononic properties. We note that TMTs crystallize in well-defined forms distinguished by (i) transition metals in trigonal prismatic coordination, usually distorted to some degree, *i.e.,* as $M^{4+}(X_2)^{2-}X^2$ with inequivalent metal–chalcogen bonding and chalcogen–chalcogen pairing; (ii) organization of these trigonal prisms into 1D chains through metal–chalcogen–metal bridging or metal-metal bonding; (iii) further organization of these 1D chains into flat or corrugated 2D sheets (bilayers) through additional metal–chalcogen bonding; and (iv) stacking of these sheets through vdW interactions. This structural hierarchy is illustrated for $TaSe_3$ in Figure 2 and can be summarized as a two-dimensional assembly of quasi-1D chains. The strength of the cross-chain M–X and X–X bonds determine whether the electronic, thermal, and vibrational properties of the TMT have the anisotropy of the 2D or 1D material [53,63–65]. Together these structural characteristics lead directly to variable band gaps, electronic instabilities like charge density waves (CDWs), and highly anisotropic optical and electrical properties advantageous for many applications. The presence of the vdW gaps facilitates both disassembly, *i.e.* exfoliation, and reassembly, *i.e.* heterostructure formation.

Conventional exfoliation techniques, which include mechanical or solvent-based sonication-assisted action, cause TMT crystals to exfoliate into ribbons or *needle*-like structures; this result is unlike that obtained with TMDs, which exfoliate into 2D sheets. In principle, such vdW materials containing 1D motifs can exfoliate completely, producing individual *atomic chains* with a cross-sectional area on the order of 1 nm × 1 nm. In contrast, materials like $Ta_2NiS_5$ and $MTe_5$ (M = Zr, Hf) incorporate 1D motifs in the form of linear chalcogen-bridged metal chains [66,67], but also include additional metal–chalcogen and chalcogen–chalcogen bonding between the 1D motifs that enhance their 2D character (see Figure 2). These effects are strong enough in $Ta_2NiS_5$ to yield 2D sheets upon exfoliation [68–70], whereas these effects are somewhat weaker in $MTe_5$, leading to nanoribbons [71–73].

It is convenient to introduce more definitive terminology about what constitutes *quasi*-1D vs. *true*-1D among vdW materials. We would consider the vdW material to be a true-1D if it contains covalent bonds only in the direction of the atomic chain, with all other bonds being of vdW type. By this criterion, $Nb_2Se_9$ and $NbCl_4$ are true-1D materials. We will call the materials quasi-1D if they contain strong covalent bonds along the 1D chain direction, while also bonded by substantially weaker covalent bonds in the perpendicular plane. In this sense, $TaSe_3$ is a quasi-1D material, as is $(TaSe_4)_2I$ due to interchain interactions *via* iodide. Although the exact number of materials with quasi-1D or true-1D structures is not known with certainty, machine learning investigations [74–76] suggest that there are hundreds of 1D vdW materials to explore. In practice, the chemical or mechanical exfoliation of 1D vdW crystals yields *bundles* of atomic chains with cross-section areas on the order of 10 nm × 10 nm to 100 nm × 100 nm, which may have nanowire or nanoribbon morphologies. If such bundles consist of only a few atomic chains, we will refer to them as atomic *threads*. We will apply the term *1D quantum materials* to those examples with quasi-1D or true-1D vdW structures on the individual atomic chain or thread scale, or when even relatively thicker bundles of such materials reveal strong quantum character. These 1D vdW materials are fundamentally different from traditional "nanowires" or "quantum wires" — high aspect ratio structures grown from conventional covalent or ionic semiconductors, *e.g.* Si or GaAs, or metals, *e.g.* Cu or Al. Such nanowires have larger diameters (~5 nm – 50 nm) and their interfaces are not necessarily atomically sharp like in vdW materials. Owing to their crystal structures, these nanowires cannot be downscaled to individual atomic chains. 1D vdW materials are also distinct from nanotubes, which can be considered as rolled-up 2D materials.

**Properties of 1D vdW Materials:** Numerous quasi-1D and 1D vdW materials, in their bulk form, have been heavily investigated since the early 1960s [77,78]. They have been mostly synthesized by the CVT method (see Figure 3). The interest in these materials peaked by the 1980s, as can be seen by the publication of several books on the topic [79–81]. Even in bulk form, many of these crystals exhibit electronic, phononic, and thermal properties derived from their 1D character, which is what has made them so interesting [82]. For example, the low dimensionality of the metallic or semi-metallic TMTs makes them susceptible to phase transitions, such as superconductivity [83–86] and CDW formation [87–90]. During the 1970s, there also was extensive work on metallic organic systems, such as TTF-TCNQ [91–93]. These molecular solids are quasi-1D metals that exhibit similar CDW phase transitions and 1D physics as the TMTs. The CDW instability of TMTs has motivated decades of research on CDW transitions [64,94,95], sliding [96–99], dynamics [100–106], dimensional scaling [107–114], and device applications [115–118]. The Fermi surface nesting, which is facilitated by quasi-1D electronic band structures, and the resulting Kohn anomaly at the CDW wavevector in the phonon spectrum of $ZrTe_3$ has been a topic of long-standing and continued focus [119–127].

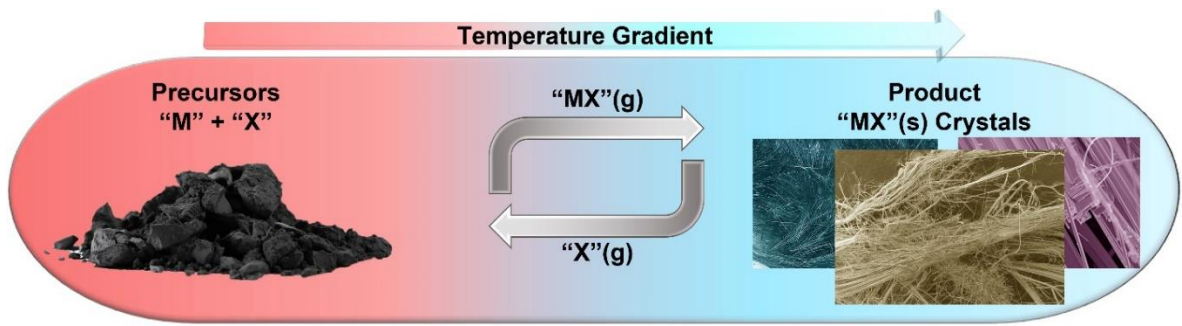

[**Figure 3:** Illustration of the chemical vapor transport method of crystal growth. Precursors may include metals or metal compounds "M" as well as chalcogens, chalcogen compounds, halides, or halide compounds "X". The gas-phase species "MX" generated at elevated temperatures circulate within the sealed reaction tube under the influence of a temperature gradient. Crystallization occurs at the growth zone.]

**Mixed-Dimensional 1D–2D Materials:** In recent years, the resurgence of interest in 1D vdW materials has evolved from work on 2D vdW materials [128]. Timely reviews of the current state-of-the-art in TMTs are provided in Refs. [53,54,129,130]. One highlights the gating of $TiS_3$ as a potential channel material for field-effect transistors (FET) [131–135] or p-n junctions [136]. The measured two and four-terminal room temperature mobilities of a BN encapsulated 26 nm thick slab of $TiS_3$ gave values of 54 and 122 $cm^2V^{-1}s^{-1}$, respectively [137]. These values are comparable to the measured mobilities of 2D TMD materials like $WS_2$ and $WSe_2$ [138] and less than the electron mobility of the 2D III-VI material InSe [139]. The anisotropies of the TMTs vary from being somewhat quasi-2D type to largely quasi-1D type. Even within the same material, different electronic bands can have opposite anisotropies, with $TiS_3$ being one example. The electron and hole effective masses of $TiS_3$ have opposite anisotropies. The calculated electron (hole) effective masses in the chain and cross-chain directions are 0.41 (0.98) $m_0$ and 1.47 (0.32) $m_0$, respectively[139], and the calculated values of the hole masses are consistent with those extracted from ARPES measurements ($0.95 \pm 0.09\ m_0$ and $0.37 \pm 0.1\ m_0$) [140]. The coupling of the electron band is strongest along the chain direction, as one would expect from the bonding. However, the coupling of the hole band is strongest along the cross-chain direction. This surprising anisotropy is the result of the different orbital compositions of the two bands [141]. The valence band is composed of predominantly S $p_x$ and Ti $d_{xy}$ orbitals, which have maximum overlap along the cross-chain direction. The conduction band composition has a large component of Ti $d_{z^2}$ orbitals, which couple most strongly along the chain direction. Thus, the anisotropy is determined not just by the geometry, but also by the orbital compositions of the bands. The band anisotropies can then give rise to similar anisotropies in response functions such as conductivity.

Another well-studied example in which the cross-chain coupling leads to a more dispersive band than the coupling along the chains is the semi-metallic TMT $ZrTe_3$. In this case, Te $p_x$ sigma bonds across the top and bottom of each layer give rise to highly dispersive bands in the cross-chain direction [120]. This results in the greatest electrical conductivity being in the cross-chain direction. The ratio of the cross-chain to through-chain electrical conductivities is 1.4 at room temperature [142]. This anisotropy between the cross-chain and through-chain conductivities is small compared to most other TMTs. Anisotropy is also present in the phonon spectra and the thermal conductivities [65]. For $ZrTe_3$, the highest velocity acoustic phonon mode is also in the cross-chain direction; however, the highest thermal conductivity is in the through-chain direction. This apparent discrepancy is a result of the fact that about 50% of the heat is carried by the optical modes at room temperature. The calculated ratios of the through-chain to cross-chain conductivities for $ZrTe_3$ are 2.5, and the ratios of the through-chain to cross-plane thermal conductivities are 4.1. Overall, $ZrTe_3$ is an example of a TMT that is closer to a quasi-2D material than a quasi-1D material. In comparison, the calculated ratios of the through-chain to cross-chain conductivities for $TaSe_3$ are 7.7, and the ratios of the through-chain to cross-plane thermal conductivities are 47, indicating that this is a TMT with greater quasi-1D nature than $ZrTe_3$.

**Tuning the Properties of Quasi-1D Materials:** The outstanding properties of quasi-1D materials can be modified using the methods of solid-state chemistry to alter the composition, control defects, and access various structures. Together with effects derived from downscaling and reduced dimensionality, these methods provide ways to engineer physical and electronic properties. For TMTs, crystalline compositions can include group 4 (Ti, Zr, Hf) and group 5 (Nb, Ta) transition metals, with some work on $MoS_3$ indicating that group 6 may be possible but amorphous [143]. Elemental substitution leads to further compositional diversity in the form of doping, involving the substitution of a relatively small number of metal or chalcogen atoms, or alloying, which ideally includes the full range of substitution, i.*e*., the complete mixing of two binary TMTs to yield ternary or quaternary compositions $M_xM'_{1-x}(X_yX'_{1-y})_3$ (where x,y≥1), also called "solid solutions." In early examples of Zr- or Ta-doped NbSe$_3$ crystals, the dopant led to increased superconducting transition temperatures up to 3 K (compared with 0.5 K in pure NbSe$_3$) by suppressing the competing CDW phases [144,145]. More recent reports have focused on the transport properties of the mixed metal alloys (Ta,Nb)S$_3$ and (Ti,Zr)S$_3$, as well as mixed chalcogen compositions Zr(Te,Se)$_3$ .[146,147] The thermoelectric, electrochemical, and catalytic properties of TMTs also are modified by substitution, as in the case of (Ti,Nb)S$_3$, while crystalline structure and anisotropy are retained [148–150]. Furthermore, the intercalation of atoms or ions within the vdW gap of TMTs is well known for Li$^+$ from electrochemical studies, and by analogy with TMDs, such intercalation can serve as another effective parameter for property engineering [151].

Structural defects are common and can make strong contributions to electronic properties. In TMTs, defects include metal or chalcogen vacancies, although undesirable atomic substitutions may also be considered defects. The effects of chalcogen vacancies have been most thoroughly studied for TiS$_3$ and ZrS$_3$. In these materials, "open" sulfur sites lead to greater reactivity, which can be positive (improved photocatalytic oxidation of molecules) or negative (increase in undesired surface oxidation) [152,153]. Additional studies showed that sulfur vacancies in TiS$_3$ reduce electron-hole recombination while promoting electron conductivity, charge separation, and transport. These properties have useful implications for photovoltaic, photoelectrochemical, and optoelectronic devices [154,155] Other TMTs like TaSe$_3$ also often exhibit chalcogen vacancies. These typically appear as non-stoichiometry from compositional analysis, *i.e*., TaSe$_{3-\delta}$. The effects of defect density and distribution in these systems are not yet clear.

Polymorphs are compositions that solidify with more than one unique arrangement of atoms (see Figure 4). The structural differences among polymorphs may be relatively small or quite significant, and can profoundly affect a material's properties. Some TMTs crystallize in polymorphic forms. In contrast to TMDs, the metal coordination in TMTs is consistently trigonal prismatic, thus structural diversity originates from a combination of (i) distinct chain packing arrangements within MX$_3$ bilayers, with varying 1D-2D character, (ii) different bilayer stacking sequences, (iii) the degree of chalcogen–chalcogen bonding, *i.e*., X$_2$ interactions in the formulation of MX$_3$ as M$^{4+}$(X$_2$)$^{2-}$(X)$^{2-}$, and (iv) the presence or absence of metal-metal bonding along the chains, which leads to metal pairing. $NbS_3$ is an especially interesting case with up to five distinct structures observed experimentally to date and several more predicted phases [30,156]. All structures contain 1D chains of stacked, niobium hexasulfide trigonal prisms assembled into bilayer sheets through interchain Nb---S and S---S interactions, similar to $TaSe_3$. The polymorphs differ in the degree of metal bonding along the chains, *i.e.,* the presence or absence of niobium pairing and the repeating pattern and alignment of the bilayer sheets. These structural features are illustrated for $NbS_3$-IV and $NbS_3$-V in Figure 4. Specifically, $NbS_3$-IV exhibits alternating long and short niobium–niobium distances whereas $NbS_3$-V exhibits uniform niobium bonding distances. These features lead to dramatically different electronic properties, with $NbS_3$-I and $NbS_3$-IV containing niobium pairing and thus semiconducting nature whereas $NbS_3$-V without niobium pairing is metallic, even exhibiting CDW transitions instabilities [95,156].

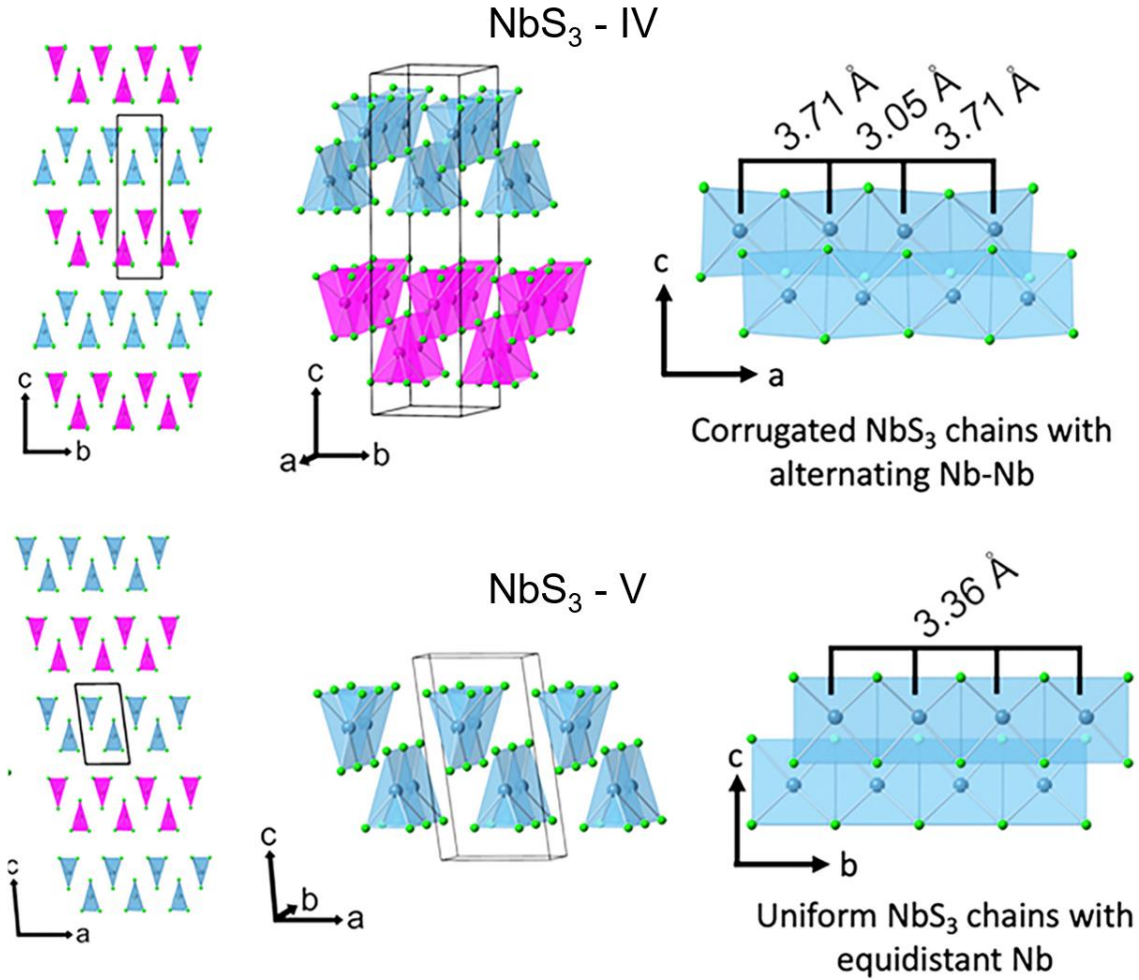

[**Figure 4:** Polymorphism of van der Waals materials. Crystal structures, unit cells, chain cross-sections, and atomic arrangements of two polymorphs of $NbS_3$-IV and $NbS_3$-V are shown from different views. The $NbS_3$-IV structure has alternating long and short Nb–Nb distances with semiconducting nature whereas the $NbS_3$-V structure has uniform Nb–Nb bond length leading to metallic and charge-density-wave properties [30].]

Recent reports indicate that strain can strongly influence the electronic structure of TMTs [157]. For instance, DFT calculations on $TaSe_3$ single layers under biaxial strain reveal metallic behavior whereas uniaxial strain along the chains induces semiconducting behavior [158]. Together, all these compositional and structural parameters can allow us to fine-tune the electronic characteristics of quasi-1D vdW materials, including the magnitude and position of the bandgap and the presence of useful electronic instabilities. The geometry of 1D materials suggests that strain can affect the properties of such materials substantially stronger than those of 2D or 3D materials.

**Breaking the Chains:** The isolation of *individual atomic planes* was the first step leading to the vast field of graphene physics, chemistry, and engineering: the Dirac dispersion and corresponding exotic properties only appeared after graphite was mechanically exfoliated into single-layer graphene [19]. However, some 2D materials reveal more interesting properties when a specific number of atomic planes are retained. For example, one needs at least five quintuple layers of $Bi_2Te_3$ to preserve the topologically protected Dirac surface states. Below this thickness, the surface states hybridize, gap, and acquire a mass [159–162]. One needs at least nine layers of 1T-$TaS_2$ to stabilize the commensurate CDW phase [163–165]. Strictly odd or even numbers of layers are required to alter the optical response of $MoS_2$ [166–168]. Drawing on this analogy with 2D materials, we can expect that the properties of 1D vdW bundles can be significantly different than those of bulk 1D vdW crystals, and we can anticipate that the properties of 1D threads will depend on the exact number and arrangement of constituent atomic chains and be distinct from those of thicker bundles. Breaking interchain interactions to generate atomic bundles unleashes interesting properties and possibilities. The 1D vdW bundles, unlike conventional nanowires of covalent or ionic semiconductors, may have smooth interfaces without dangling bonds. In contrast to nanowires of elemental metals, which are always polycrystalline, metallic 1D vdW bundles can be single crystalline, creating opportunities for the suppression of electromigration.

The range of materials with 1D motifs, which can be potentially broken down to individual atomic chains or bundles of atomic chains, is rather broad. The data mining and machine learning studies used algorithms that determined the dimensionality of weakly bonded crystals by considering the positions of atoms in bulk 3D crystals. The application of this algorithm to the Materials Project database of over 50,000 inorganic crystals identified 487 materials containing 1D chains weakly bonded to neighboring atoms [74,169]. These materials include semiconductors, metals, and electrical insulators. One of the first studies identified more than 40 quasi-1D materials with a bandgap $E_G$=0 eV, more than 10 with $E_G$= 1.5 eV, and around 10 with $E_G$=2.5 eV. The vdW interactions and atomically sharp interfaces enable convenient hetero-integration, including vertical and horizontal stacking and the possibility of vdW epitaxy. Theoretical studies indicate that the quantum confinement of charged carriers in 1D vdW materials manifests differently than that in nanowires of covalently bonded solids: it often does not open the band-gap, meaning that electrons in one atomic chain are not strongly affected by the absence of neighboring chains. Most of the 1D vdW materials studied to date have been obtained by the "top-down" approach: first, relatively large (mm length), high quality, single crystals are synthesized by CVT or flux growth methods, and then these are exfoliated either chemically or mechanically. In recent work, the "bottom-up" CVD of quasi-1D vdW materials has been implemented with some control over bundle size, position, and orientation [31]. With bundles in hand, the fabrication of test structures for measuring properties of the individual quasi-1D vdW channels typically follows similar techniques as in 2D materials research, using electron beam lithography or shadow masks.

**The Ultimate Limit of Downscaling:** Can we reach the ultimate limit of individual 1D vdW atomic chains? We note that the Mermin-Wagner theorem, which states that long-range order cannot exist in ideal 1D or 2D structures, does not prohibit real-world 1D vdW atomic chains. In practice, all physical structures, 2D or 1D, have 3D degrees of freedom. Consider extremely stable 2D graphene: the fact that it does have 3D degrees of freedom is revealed by out-of-plane crystal lattice vibrations, termed acoustic *ZA* and optical *ZO* phonon modes. These phonon modes, among other effects, limit the phonon thermal conductivity of ideal graphene. Thus, fundamental physics laws do not prohibit reaching the limit of downscaling. Reports of 1D vdW atomic chains stabilized within carbon or boron nitride nanotubes provide promising precedent for their isolation [170,171]. Placing 1D materials on substrates can further increase their stability in the same way as with 2D materials. Technologically, achieving the atomic-chain limit with both true-1D and quasi-1D crystals is feasible by chemical or mechanical exfoliation. The *ab-initio* theoretical data indicate that the out-of-plane, *i.e.* along the c axis, cleavage energy of quasi-1D vdW crystals is close to or even lower than that of graphite, measured to be ~0.37 $Jm^{-2}$ [172,173]. For example, the cleavage energy for $TiS_3$ is ~0.226 $Jm^{-2}$. The in-plane cleavage energy between the chains is slightly greater than that of the out-of-plane direction. For $TiS_3$, the value is ~0.5 $Jm^{-2}$. However, along the chains, the cleavage energy is significantly greater, on the order of ~3 $Jm^{-2}$ [174]. For these reasons, it should be possible to identify the appropriate experimental parameters for the exfoliation of individual chains even if there is a difference in the in-plane and cross-plane binding energies.

**The Quantum States on a Grand Scale:** The definition of *quantum* varies depending on the discipline, research field, and, sometimes, the funding agency. The 1D vdW materials described here are genuinely quantum even when spatial confinement does not result in the band-gap opening. Many 1D vdW materials are strongly-correlated systems. They reveal electronic correlations linked to non-trivial quantum phenomena [33,34]. Their properties are governed by collective quantum behavior, which results in the emergence of CDWs and superconductivity, or the formation of excitons and polaritons [83,175]. Historically, the first reports of the properties of quasi-1D vdW materials appeared specifically in the context of CDW research [103,107,114,176]. It happened to be that the classical CDW materials, *e.g.* $NbS_3$ and $TaS_3$, are vdW metals with 1D motifs as the building blocks of their crystal structures. One important consequence is that many vdW materials reveal multiple CDW phases, which are defined as the *macroscopic quantum states* or *quantum condensates* with the period equal to half the de Broglie wavelength of an electron at the Fermi level, $\lambda = \pi/k_F$ ($k_F$ is the Fermi wave vector) [90,100,102]. This quantum state formation is responsible for the opening of the CDW energy gap in the electronic spectrum at the Fermi level of these material systems, which otherwise are normal metals (see Figure 5).

Initially, CDW properties were studied in bulk crystals [88,89,116,177–179]. The first reports of vdW crystal thinning to understand size effects and attain the regime of just a few conducting atomic channels started appearing around 2000 [55,95,103,104,107,176,180–186]. The main drivers were to obtain coherent CDW transport regimes, which are easier in a few channels than in the bulk, and to inhibit single-particle excitations while retaining collective CDWs. It was also noted that, in the absence of the long-range interaction in 1D metallic systems, one should expect the formation of a Luttinger liquid and, at small concentrations of charge carriers, the formation of a one-dimensional Wigner crystal. The results for bundles, *i.e.* nanowires, of $NbS_3$, $TaS_3$, and other CDW materials have been exciting in terms of physics and possible applications. It has been suggested that decreasing the cross-sectional area of bundles of the atomic chains of quasi-1D materials can improve the coherency of the collective CDW transport and reveal intriguing phenomena associated with the single quantum phase-slip events, *i.e.*, change in the number of CDW periods $N$ by $\pm 1$. If the metal contacts create stationary boundary conditions for the CDW phase, $N$ is an integer and the electron wave vector $q=2\pi N/L$ adopts discrete, *i.e.* quantum, values. Such quantization of $q$ can be viewed as one quantum effect superimposed on another quantum effect—with the CDW already constituting a quantum condensate with the period equal to half the Fermi wavelength. The extra quantization—the discrete states of CDWs—is similar to the quantum confinement effects for electrons in a potential quantum well. The conclusions one draws from the pioneering studies of quasi-1D van der Waals materials like $NbS_3$ is that even the bundles of atomic chains with the cross-sectional dimensions on the order of 100 nm × 100 nm are quantum materials revealing collective quantum states and associated unique transport properties, even at RT [95,104,180].

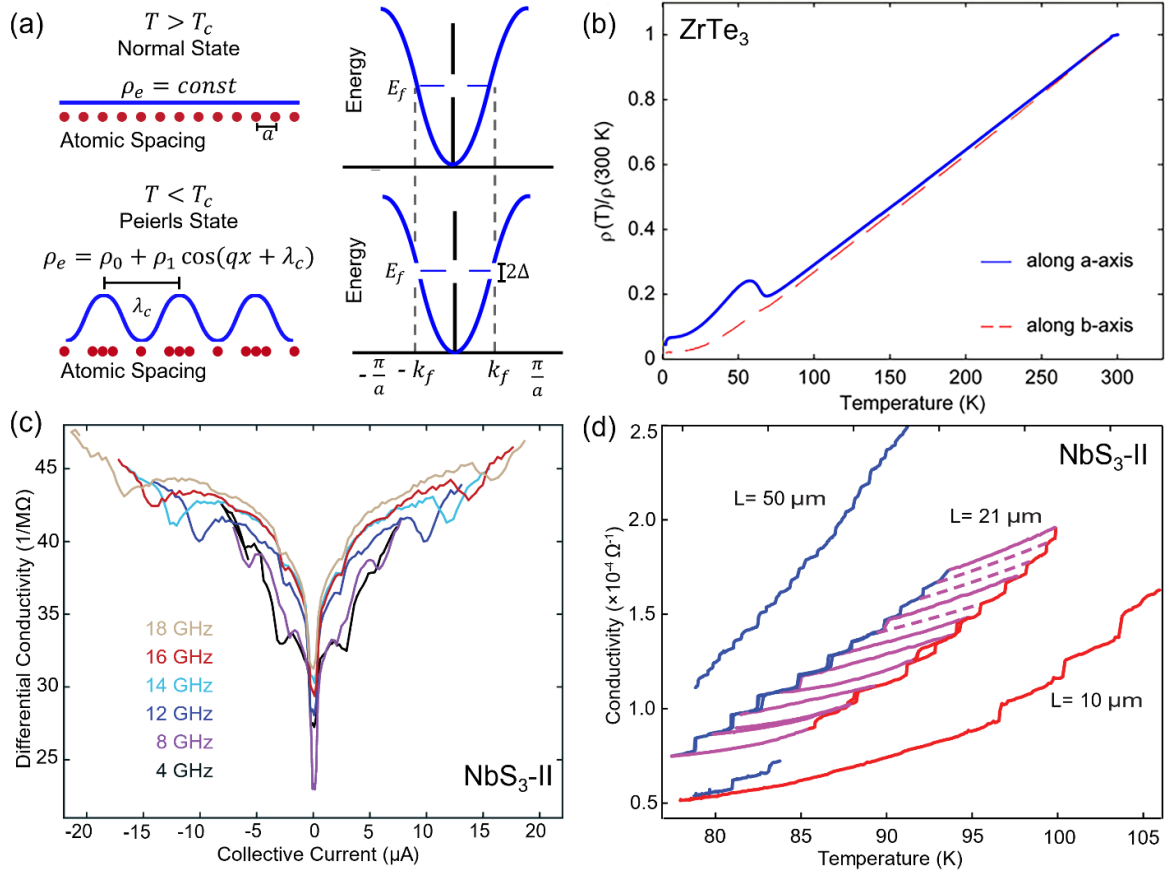

[**Figure 5:** Physics of charge-density-wave quantum states. (a) Schematic showing the Peierls distortion in a one-dimensional metal. Note the reordering of electrons and change in the charge distribution below the CDW phase transition temperature, $T_c$. The distortion opens up a gap at the Fermi level as shown in the electron energy diagram on the right. (b) The normalized resistivity of quasi-1D ZrTe$_3$ along (b-axis) and perpendicular (a-axis) to the chains as a function of temperature. Note the clear resistivity anomaly at $T$~63 K, indicating the CDW transition [123]. (c) The differential conductivity of NbS$_3$-II as a function of the collective current at room temperature under RF radiations with different frequencies. The Shapiro steps are observed at frequencies as high as 20 GHz [95]. (d) The temperature-dependent electrical conductivity and hysteresis loop in K$_{0.3}$MoO$_3$ for three samples with different lengths, $L$. The arrows indicate the temperature scan direction. The steps observed for the sample with $L$ = 21 μm, show the quantization of charge-density waves [55].]

**Quasi-1D Topological Insulators:** At present, the study of topological phases of materials is one of the most active research areas in solid-state physics, with a plethora of topologically nontrivial quantum phases being predicted theoretically and discovered experimentally [46,187–189]. Topological insulators are electronic materials that have a bulk bandgap but reveal protected conducting states on their edge or surface. These states are possible due to the combination of the spin-orbit interactions and time-reversal symmetry. In topological insulators, the non-trivial topological character of bulk electron wave functions guarantees the presence of the gap-less metallic surface states. The energy-momentum dispersion of surface states acquires a conical shape referred to as the Dirac cone dispersion. Until recently, the investigated topological insulators materials were either 3D strongly-bonded bulk materials or quasi-2D vdW materials, such as compounds of the $Bi_2Se_3$ family. The 3D topological insulators are classified as either strong or weak, and most experimental reports focus on the strong topological insulators. The weak topological insulators are more challenging to confirm experimentally because the topological surface states emerge only on particular side surfaces, which are typically undetectable in real 3D crystals. Recently, the 1D topological insulators were predicted and discovered in quasi-1D vdW crystals of β-$Bi_4I_4$ [47,188]. Owing to the quasi-1D structure of this material, the surface Dirac cone exhibits strong anisotropy: the velocity of surface electrons moving parallel to the atomic chain direction is five times larger than that of surface electrons moving perpendicular to it, which is in striking contrast to 3D or 2D vdW topological insulators. The high anisotropy of the surface-state Dirac fermions suggests the possibility of combining topological order with other types of ordering characteristics to one-dimensional systems, *e.g.* CDW phase discussed above. One can expect that the topological quantum phase in quasi-1D materials can be controlled by applying proper external strain or bias, particularly as the material is thinned down to bundles of atomic chains.

There have been recent developments pertinent to the quantum quasi-1D vdW materials where topological phase transitions emerge as a result of modifications of the single-particle band structure due to the strongly correlated interactions. The best-known example is monoclinic $(TaSe_4)_2I$, which consists of $TaSe_4$ chains with helical symmetry that are separated by chains of iodine atoms [190]. This material is a Weyl semimetal with numerous Weyl points located above and below the Fermi level, forming pairs with the opposite chiral charge (see Figure 5). Simultaneously, the quasi-1D $(TaSe_4)_2I$ reveals the incommensurate CDW phase forming at temperatures below the Peierls transition temperature $T_P$=248 K - 263 K. It has been recently demonstrated that the CDW phase in $(TaSe_4)_2I$ couples the bulk Weyl points and opens a bandgap. This correlation-driven topological phase transition in $(TaSe_4)_2I$ provides a route towards observing the solid-state realizations of axion electrodynamics in the gapped regime, as well as topological chiral response effects in the semi-metallic phase [191,192]. It has been suggested that $(TaSe_4)_2I$ reveals a correlated topological phase, which arises from the formation of CDW in a Weyl semimetal. This quasi-1D quantum material truly represents an exciting avenue for exploring the interplay of correlations and topology in solid-state systems.

**Carrying Current *via* Atomic Threads:** Quasi-1D vdW materials have revealed interesting properties related to their electrical current-carrying capabilities. It was discovered that the exfoliated metallic TMT bundles with the cross-section dimensions in the range from 10 nm to 100 nm reveal exceptionally high breakdown current densities – on the order of 10 MA/cm$^2$ for $TaSe_3$ [26], and approaching 100 MA/cm$^2$ for $ZrTe_3$ (see Figure 6) [28]. For comparison, the breakdown current density for the technologically important metal such as Cu is ~3 MA/cm$^2$ [193,194]. Interestingly, in some cases, the breakdown in quasi-1D vdW metals was step-like when weakly bound atomic chains or atomic threads break one by one. There was also an indication of self-healing when the increased current through the surviving atomic threads results in the increase in local temperature followed by the reconnection of the neighboring atomic threads [26]. The high current densities sustained by quasi-1D vdW materials can be attributed to several characteristics of such materials. First, they are *single crystals* in one direction and have few or no dangling bonds. The vdW interfaces can be nearly perfect. In contrast, elemental metals such as Cu or Al are polycrystalline and, as a result, have grain boundaries and natural interface roughness. The grain boundary diffusion, which leads to electromigration damage, has been identified as the dominant breakdown process in Al interconnects whereas surface diffusion in Cu interconnects. Since the current breakdown in metals is of the electromigration rather than thermal nature one can conclude that the electromigration activation energies and the on-set of breakdown can happen in the single-crystal 1D vdW materials at substantially higher current density [27,29]. The nearly perfect, low roughness, of the interface of quasi-1D vdW materials may also reduce the thermal boundary resistance between the 1D channel and the substrate, resulting in better heat removal and higher sustained current levels. There is an intriguing connection between the high current densities achievable in quasi-1D vdW metals and their ability to reveal the quantum effects. The CDW velocity and fundamental frequency of the quantum CDW condensate oscillation are proportional to the current density. The record-high CDW oscillation frequencies of ~200 GHz were achieved in NbS$_3$ whiskers at the current density of ~6 MA/cm$^2$ [55]. It is anticipated that the reduction in the cross-section area of such quasi-1D vdW conductors will result in even higher frequencies.

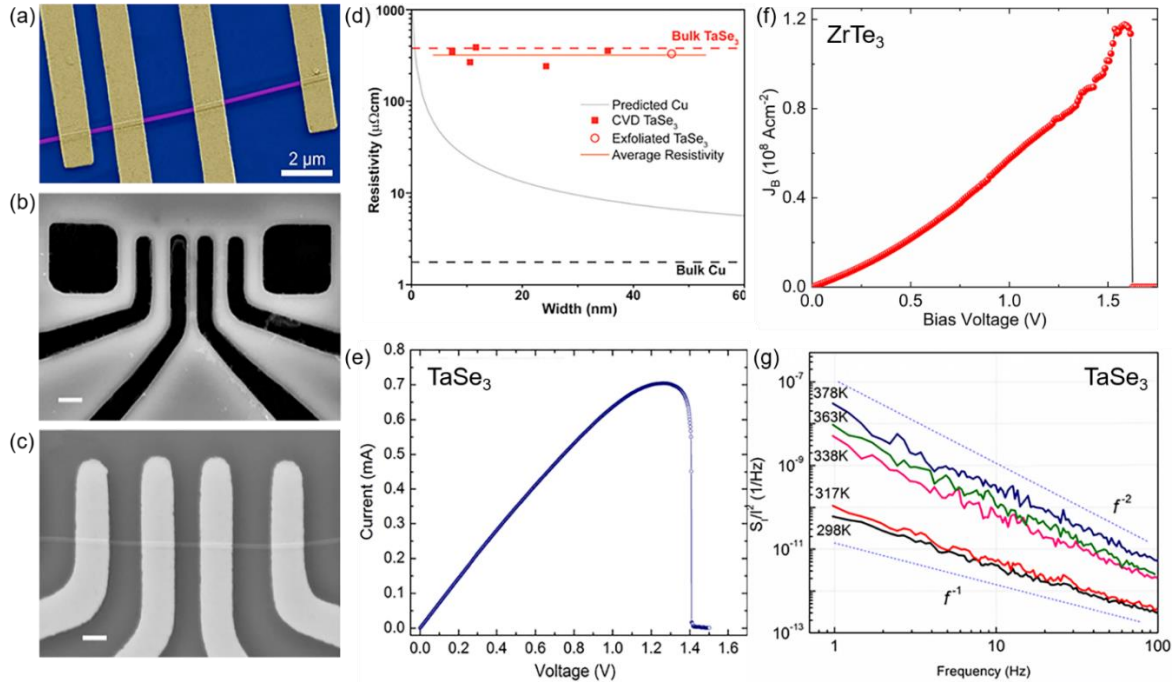

[**Figure 6:** Current density in quasi-1D van der Waals metals. The SEM images of (a) a representative test structure fabricated with quasi-1D TaSe$_3$ nanowires using EBL [26], (b) a shadow mask with TLM structure patterned on 500-µm thick Si/SiO$_2$ wafer [28], and (c) a quasi-1D ZrTe$_3$ device, fabricated by the shadow mask method. The four-contacts were used for measuring I-Vs, contact resistivity, and current density. The scale bars in (b) and (c) are 2 µm and 1 µm, respectively. (d) Resistivity of CVD-grown and mechanically exfoliated TaSe$_3$ nanowires as a function of resistivity-sectional dimension. The bulk resistivity values of TaSe$_3$ and copper (dashed lines) as well as a prediction for the scaling of the copper resistivity with wire width (solid line) are shown for reference. Unlike in elemental metal nanowires, the resistivity of quasi-1D metallic nanowire does not increase with decreasing cross-section [31]. (e-f) High-field I-V characteristics of quasi-1D nanowires of TaSe$_3$ [26] and ZrTe$_3$ [28]. The breakdown current density values were reported to be up to 10 MA/cm$^2$ and 100 MA/cm$^2$ for TaSe$_3$ and ZrTe$_3$, respectively. (g) The normalized current spectral density, $S_I/I^2$, measured for qusi-1D TaSe$_3$ nanowires at different temperatures. The $1/f$ noise becomes more of $1/f^2$ at elevated temperatures [27]. ]

The disadvantage of experimentally tested quasi-1D vdW metals so far has been their substantially higher bulk resistivity as compared to that of Cu. However, there are recent reports that indicate that the resistivity of quasi-1D vdW conductors does not increase much or even stays constant with the decreasing cross-sectional area of the conductor [31]. The latter is in the stark contrast to Cu and other elemental metals where the resistivity rapidly increase due to the electron scatterings on grains and nanowire boundaries as the cross-plane dimensions approach a few-nanometer range. The electrical resistivity of Cu nanowires with the cross-sectional dimensions below 10 nm can increase by more than two orders of magnitude compared to its bulk value. Not everything is defined by the grain boundaries and surface roughness. Theory suggests that quantum confinement works differently in the atomic threads of 1D vdW materials than in conventional covalently bonded quantum wires [31]. The calculations performed for $TaSe_3$ and a few other vdW materials show that the electron density of states (DOS) near the Fermi energy, $N(E_F)$, attains its maximum when the number of individual atomic chains is reduced to one – the ultimate limit. In the simple formulation, the electrical conductivity, $\sigma$, is proportional to the scattering time, $\tau(E_F)$, and $N(E_F)$, suggesting that even if $\tau(E_F)$ scales inversely with DOS, there is a possibility that the increasing $N(E_F)$ can offset this effect, and maintain the high electrical conductivity. Given this difference in the quantum confinement effects of downscaling, the absence of the grain boundaries, and the atomically sharp interfaces make the quasi-1D vdW metals potential candidates for interconnect applications in the future nanoelectronic circuits.

**Thermal Conductivity in 1D Limits:** Thermal conductivity of 1D quantum materials is an intriguing fundamental science question. The opinions on what happens to materials' ability to conduct heat, when scaled down to two or one dimensions, range from strong suppression to potentially reaching infinity [21,195,196]. It has long been known theoretically that the intrinsic thermal conductivity, $K$, limited by the crystal inharmonicity alone, is finite in three-dimensional (3D) materials. However, it reveals a logarithmic divergence in 2D crystals, $K \sim \ln(N)$, and power-law divergence in 1D systems, $K \sim N^{\alpha}$, with the system size $N$ (where $N$ is the number of atoms; $0 < \alpha < 1$) [197–201]. The divergence in thermal conductivity of 1D materials is known as the Fermi-Pasta-Ulam-Tsingou paradox which was proposed in 1955 [202]. Actual material systems are not exactly 2D or 1D and have finite thermal conductivity because of cross-plane phonon modes. Graphene is as close as one can get to 2D system; and the true-1D vdW material exfoliated to an individual atomic chain is as close as one can get to 1D system (note that SWCNTs are rather rolled up 2D systems). The experimental and theoretical evidence available now indicate that graphene's heat conduction indeed shows anomalies related to its 2D nature [203–205]. The effects for 1D systems can be even more pronounced if one carries out experiments with a suspended 1D quantum material where the transverse acoustic phonons are suppressed. Recent years witnessed other intriguing developments indicating the importance of phonon *hydrodynamic* transport [205–214]. Phonons in the hydrodynamic regime can form packets that change the typical diffusive behavior of heat and make it propagate as a wave, with the corresponding phenomenon of the second sound. In bulk materials, the phonon hydrodynamic transport regime occurs only at low temperature when the normal phonon-phonon scattering dominates the Umklapp scattering [205–214]. There are indications that the phonon hydrodynamic conduction can occur in graphene at relatively high temperature owing to its 2D nature [206,208,213]. The 1D quantum materials can be even more suitable for the hydrodynamic phonon regime due to stronger Umklapp-scattering suppression. From the other side, one can envision mechanisms for the suppression of the thermal conductivity of quasi-1D vdW materials, which go beyond the usual phonon – boundary scattering. Confinement and phonon folding in 1D quantum materials can result in hybridization of all phonon modes [215] and an increase in the Umklapp scattering. The actual effect will likely depend on the specifics of the 1D crystal structure [216–223]. The suppression of the thermal conductivity will be beneficial for thermoelectric applications [221,224–226].

The first reports on the thermal conductivity of quasi-1D materials have started to appear [227,228]. been measured experimentally using various techniques. The thermal conductivity of $(TaSe_4)_2I$ in the chain and perpendicular to chain directions has been measured using a four-contact method [227,228]. The thermal conductivity reveals an anomalous temperature behavior close to the CDW phase transition temperature $T_P$ [227,228]. The thermal conductivity data exhibits a well-defined minimum below $T_p$ followed by a cusp at $T_p$. This anomalous behavior is explained by the contribution of thermally-assisted phason motions in addition to the main heat carriers, *i.e.*, phonons and electrons [227–229]. Electrical and thermal conductivities of quasi-1D $NbSe_3$ were measured concurrently using the microthermal bridge method (see Figure 6) [35,230]. The contribution of electrons, $K_e$, to the total thermal conductivity, $K_t$, were estimated using the Wiedemann-Frantz law. The phonon thermal conductivity was calculated as $K_{ph} = K_t - K_e$. The results presented in Figure 6 show that while $K_e$ decreases significantly at the two phase transition temperatures of 59 K and 145 K, $K_{ph}$ reveals a kink at these temperatures [227,228]. This was attributed to the strong electron-phonon interaction below the CDW transition temperatures. In a more recent study, the thermal conductivity of NbSe$_3$ as a function of nanowire length, $L$, and hydraulic diameter, $D_h$, was studied. $D_h$ is defined as the four times of the reciprocal of the surface area-to-volume ratio [35]. The length-dependent thermal conductivity for a nanowire with cross-sections of $D_h$ in the range of 10.6 to 12.3 nm, close to 1D limit, indicates that in the length range of 6.5 µm to 42.5 µm, the thermal conductivity follows a trend of $K \propto L^{\beta}$, with $\beta = 1/3$. More interestingly, the data on $K$ as a function of $D_h$ shows that $K$ first decreases with shrinking the cross-section from 135 nm to the critical cross-section of $D_h = 26$ nm and then starts to increase sharply with further decreasing of $D_h$. The divergence in $K$ is attributed to the transition from 3D phonon to 1D phonon transport. These results shed light on the Fermi-Pasta-Ulam-Tsingou paradox and reveals a possibility of finding thermally superconducting materials among the members of 1D vdW crystals.

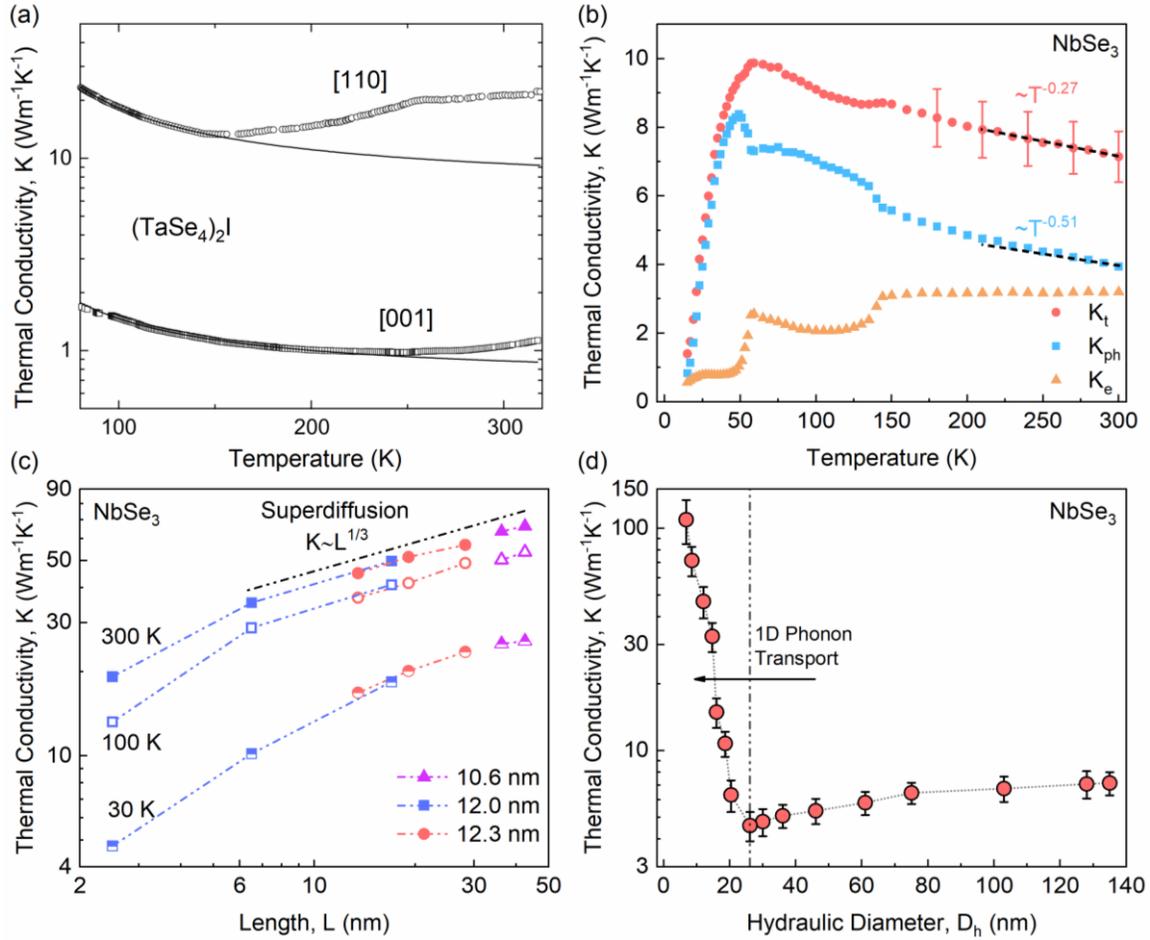

[**Figure 7:** Thermal conductivity of 1D van der Waals materials. (a) Thermal conductivity of bulk $(TaSe_4)_2I$ along ([100]) and perpendicular to the chain ([100]) directions [228]. (b) Thermal conductivity of charge-density-wave $NbSe_3$ nanowire. The cross-section of the sample is 135 nm. In the plot, $K_T$, $K_e$, and $K_{ph}$ represent the total, electron, and phonon thermal conductivity, respectively. Note the variation of $K_e$ and $K_{ph}$ at the CDW transition temperatures [230]. (c) The measured thermal conductivity of $NbSe_3$ as a function of the nanowire length, $L$, at different temperatures with similar hydraulic diameters, $D_h$. Note that the thermal conductivity at 100 K and 300 K follows $L^{1/3}$ power-law divergence as a result of the phonon super-diffusion, the dominant mechanism in the 1D limit [35]. (d) The measured thermal conductivity of $NbSe_3$ nanowires with suspended length of 15 μm as a function of hydraulic diameter, $D_h$. The data exhibits a clear transition at $D_h = 26$ nm [35] where the 1D phonon transport dominates the thermal transport.]

**Nanocomposites of Atomic Threads:** Low-dimensional materials like carbon nanotubes and graphene have shown promise as fillers in various composites. Composites have a potential of becoming the first practical applications of low-dimensional materials as fillers. Carbon nanotubes are primarily used to reinforce mechanical properties or increase the electrical conductivity of polymer composites. These functionalities require relatively high filler loadings to create the necessary percolated networks. Given the high cost of carbon nanotubes, such applications are somewhat limited. Liquid phase exfoliated (LPE) graphene has proved to be an effective filler in composites for thermal management and electromagnetic interference (EMI) shielding [231–234]. Graphene couples better to the polymer matrix than carbon nanotubes and can be mass produced inexpensively by LPE [231–237]. In comparison, the 1D vdW materials offer a much greater variety of properties for composite applications than carbon nanotubes or graphene. The library of 1D vdW materials include metals and semiconductors, as well as magnetic and topological insulator materials. Many of them can be obtained in atomic bundles or thinner threads via LPE or other exfoliation techniques on scales required for composite preparation. The fact that metallic 1D vdW remain metallic when downscaled to atomic chains enables fillers with unprecedented aspect ratio, reaching $10^6$, *i.e.* millimeter length scale and nanometer diameter (see Figure 8) [42,45].

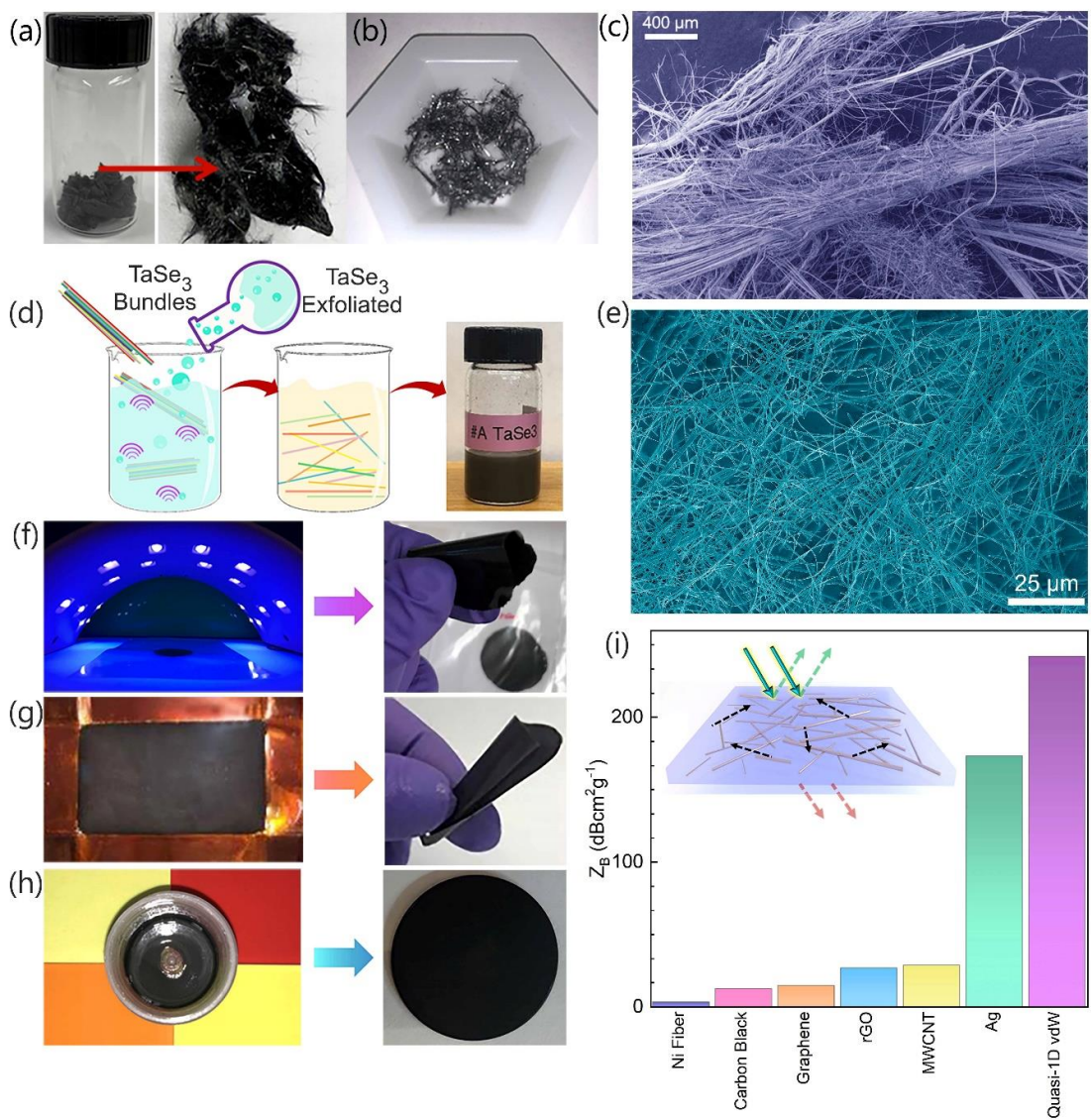

[**Figure 8:** Composites with quasi-1D van der Waals materials. (a - b) Optical images of as-prepared TaSe₃ crystals. (c) SEM image of the needle-like fibrous TaSe₃ bundles; (d) schematic showing the liquid phase-exfoliation process of TaSe₃ in solvent. The vial shows exfoliated TaSe₃ in acetone. (e) SEM image of the liquid-phase exfoliated TaSe₃ in acetone. The exfoliated motifs have high aspect ratio morphology. (f - g) Flexible films prepared with special off-the-shelf UV-cured polymer and sodium alginate with quasi-1D exfoliated TaSe₃ as fillers; (h) epoxy composite with TaSe₃ as fillers; (i) EMI shielding characteristics of polymers with different fillers. The $Z_B$ factor indicates composite's shielding effectiveness per aerial density of the filler. Composites with higher lower thickness, density, and filler weight loading fraction reveal higher shielding effectiveness characterized by the figure-of-merit $Z_B$ values [45].]

The first reports of composites with exfoliated quasi-1D fillers used TaSe₃ atomic bundles with the average cross-sectional dimensions in the range from ~50 nm to ~100 nm [42,45]. These studies demonstrated extremely high EMI shielding efficiency in the X-band (8.2 GHz – 12.4 GHz) and sub-THz EHF-band (220 GHz to 320 GHz), important EM frequency ranges for future communications technologies. Shielding of up to 70 dB was achieved with thin films (< 100 μm) at low loadings of exfoliated $TaSe_3$ (<3 vol. %) while these composites remained electrically insulating. The fact that composites with filler concentrations below the percolation threshold can interact with EM waves so strongly, and thereby deliver unique functionality, can be attributed to high current densities sustained by the quasi-1D vdW filler with its unusually high aspect ratio. These characteristics allow the $TaSe_3$ bundles to function, in a sense, as *atomic scale antennas*. Further alignment of the 1D vdW fillers can provide anisotropic composite properties previously unattainable, specifically EM polarization selectivity that mimics at the nm-scale the function of metal grid antennas [42]. One can envision more exotic properties of the composites when the quasi-1D fillers are thinned down to individual atomic chains and few-chain threads. The functionality of such composites, which the filler loading below the percolation threshold, can be defined by the intrinsic quantum properties of the fillers, *i.e.* CDW phases, which can be controlled externally by electrical field or temperature.

**Printing with Inks of quasi-1D Materials:** Printing has been introduced as an efficient method for fabrication of electronics on both rigid and flexible substrates [32,238–240]. Printing offers high volume production of electronics at lower costs compared to the conventional cleanroom nanofabrication methods. The typical ingredients of inks used in printing include metallic nanoparticles such as Ag, Au, Cu [241–244], or carbon allotropes, *e.g.*, graphene and CNTs [245,246] dispersed in suitable solvents. Most recently, functional inks with exfoliated TMDs, *e.g.*, $MoS_2$, $MoSe_2$, and $WS_2$ [247–250] have been used for printing FETs, sensors, and optoelectronic devices [238,240,250–254]. TMDs have low cleavage energy, can be easily exfoliated into quasi-2D fillers via LPE, and dispersed in ethanol, acetone, or other solvents to prepare functional inks. Quasi-1D TMTs offer the same benefits as TMDs with one specific advantage: they exfoliate into needle-like structures with extremely high aspect ratios. The latter offers a possibility of a smaller loading fraction of material required for achieving electrical conductivity [255,256]. A recent study has reported the application of inks prepared with quasi-1D $TiS_3$ dispersed in the mixture of ethanol and ethylene glycol for the printing of two-terminal electronic devices. Abrupt changes observed in I-V and electronic noise results close to TiS₃'s phase transition temperature suggest that the material most likely preserves its intrinsic CDW and metal-insulator transition properties even after LPE, mixing with solvents, and printing. These results suggest a possibility of printed electronics with diverse functionalities based on the quantum properties of TMD and TMTs [32].

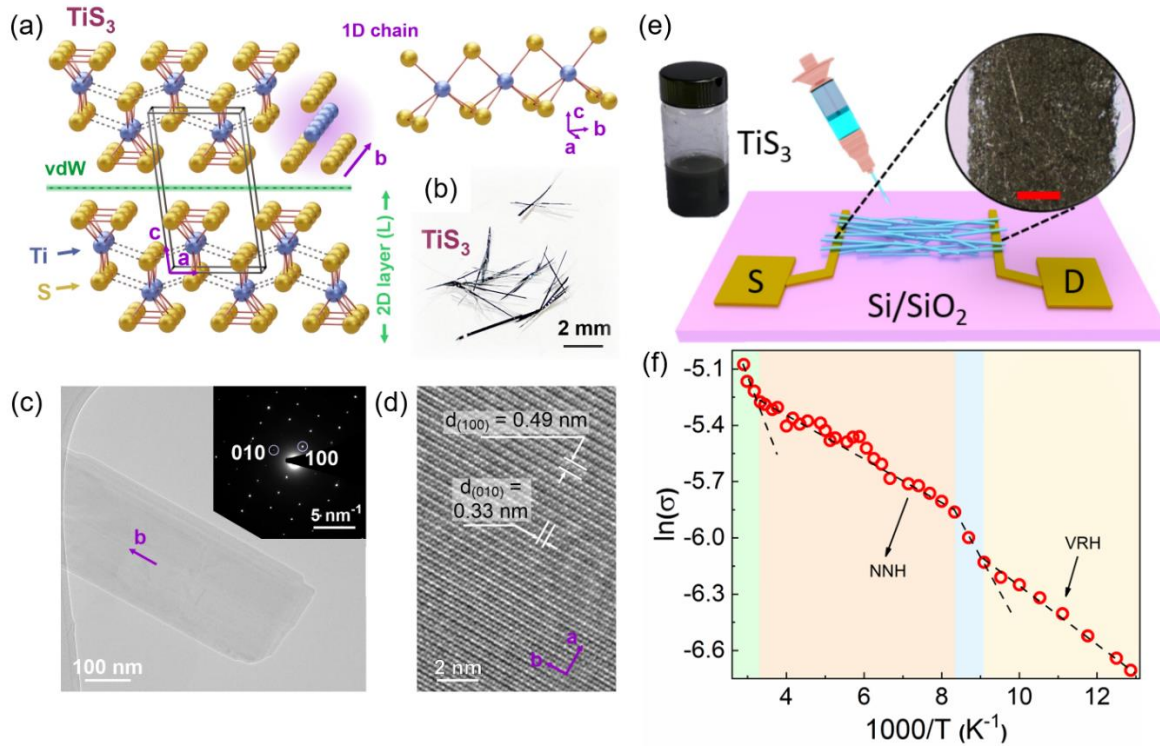

[**Figure 9**: Printing of with the inks of quasi-1D van der Waals materials. (a) Schematic showing the monoclinic crystal structure of TiS₃ from different views. (b) Optical image of the synthesized TiS₃ crystals; mote their needle-like structure. (c-d) High-resolution TEM images of the liquid-phase exfoliated TiS₃. The inset in (c) shows the selected area electron diffraction (SAED) pattern on the same crystal. (e) Schematic showing the printing process of functional inks with quasi-1D TiS₃ as filler on top of gold contacts fabricated by EBL. An optical microscopy image of the printed device is shown inside the circle. The scale bar is 200 µm. (f) Arrhenius plot of the electrical conductivity of the printed device as a function of temperature. The electrical conductivity is dominated by electron hopping mechanism. Different colors shows the change of charge transport from nearest-neighbor hopping (NNH) to variable-range hopping (VRH) [32].]

**Predicting the Properties of Quasi-1D Materials:** 487 1D materials with bandgaps ranging from metallic to over 5 eV (see Figure 10) were identified by data mining the Materials Project [74]. In this work, the algorithm based on bond lengths identified a number of the TMTs as 2D materials rather than 1D materials. The intercalated 1D materials, such as $(TaSe_4)_2I$, that have shown interesting topological properties, are missing from the 1D materials database of Ref. [74]. Therefore, this compilation of 1D materials should be treated as a good starting point for further verification and analysis. There have been a number of relatively recent works theoretically investigating the properties of monolayers of the TMTs [140,152,172,257,258]. These include studies of band alignments of Zr and Hf based TMTs with Tc and Re based TMDs for solar applications [258] and band alignments of Ti, Hf, and Zr based TMTs [152]. There have been few works investigating few-chain-bundles or single-chains of 1D materials. The database of 1D bulk materials compiled in Ref. [172] was used as the starting point for a high-throughput DFT investigation of single chains of 367 of these materials [76]. The resulting structural, electronic, and magnetic properties are provided both in machine accessible JavaScript Object Notation (JSON) files and human-readable files. Of these 367 single-chain structures, 8 were metallic ferromagnets (FMs), 14 were semiconducting FMs, 5 were metallic antiferromagnets (AFMs), and 34 were semiconducting AFMs. A recent high throughput investigation of the properties of single chains of 208 transition metal di-halides and tri-halides found a wide range of ferromagnetic, antiferromagnetic, normal metal, semiconducting, and insulating properties summarized in Figure 10 [75].

We are aware of only one theoretical investigation of few-bundle 1D materials [31]. Density functional theory was used to model the electronic structure of $TaSe_3$ in various forms: bulk, a 2D bilayer of wires, bundles of 2×2 and 2×1 wires, as well as a single wire. The resulting density of states (DOS) near the Fermi level is shown in Figure 10. In each case there is considerable DOS near the Fermi level ($E_F$) and no band gap. Furthermore, as the wire bundle is thinned to a bilayer, a finite bundle of wires, and finally a single wire, the DOS near $E_F$ increased. This is consistent with results that we have observed for 2D materials [259,260]. Considering the expression for conductivity, $\sigma_z = \frac{e^2}{3}\tau(E_F)v^2(E_F)N(E_F)$ where $\tau$, $v$, and $N$, are the scattering time, velocity, and density of states at the Fermi level, this finding suggests that conductivity can be maximized by scaling the cross-sectional area of a 1D material down to a single wire.

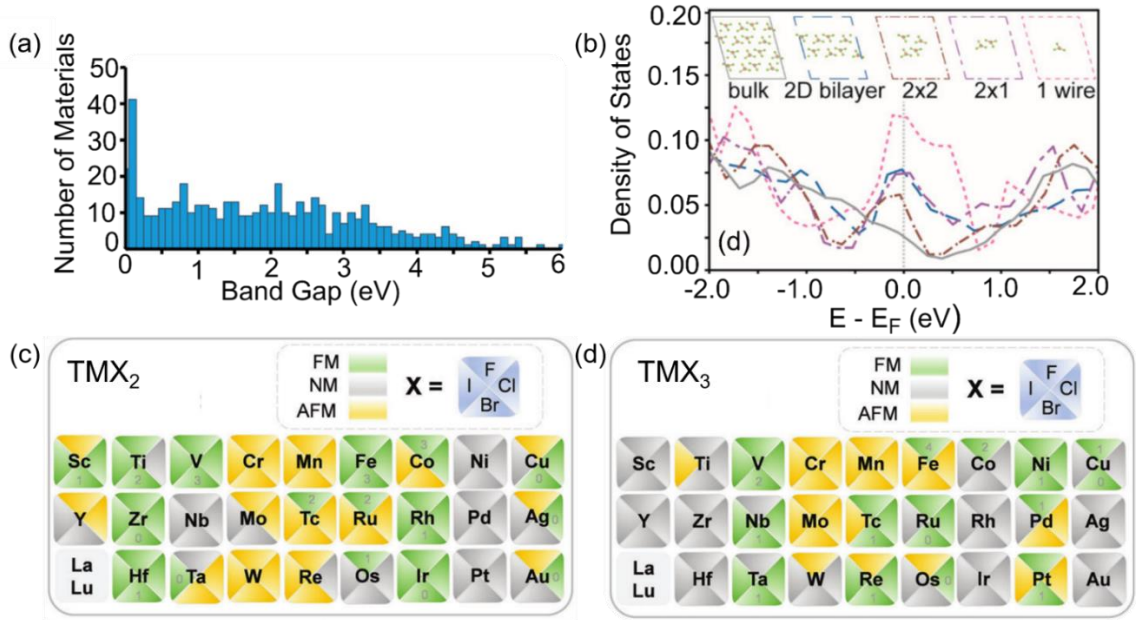

[**Figure 10:** Computational prediction of properties of 1D van der Waals materials. (a) The band-gap distribution of 1D materials identified from the Materials Project database [74]. (b) Change in the density of states for different dimensionalities and bundle sizes of TaS$_3$ [31]. Predicted magnetic properties of the transition metal (b) di-halide and (c) tri-halide wires [75].]

**Outlook:** The 1D quantum materials based on exfoliated or grown atomic bundles and threads of vdW crystals with 1D motifs have already demonstrated properties that distinguish them as a unique class of low-dimensional materials. The quantum nature of the quasi-1D vdW crystals, associated with the strongly correlated CDW condensate and topological insulator phases, can reveal itself even in the bulk samples. Exfoliating to atomic bundles with cross-sectional dimensions on the order of 10 nm to 100 nm enables increased coherence of CDW phases, and it can lead to intriguing phenomena, such as quantization of the CDW condensate and the electrical conductivity steps associated with single quantum phase-slip events. It is exciting that these quantum effects can be observed at room temperature and in a sample a few-μm in length. The quasi-1D vdW atomic bundles have shown promise as exceptional current conductors and fillers in functional composites. Their atomically sharp vdW interfaces and single-crystalline nature support current conduction at ever decreasing cross-sections whereas conventional polycrystalline elemental metals experience a fast increase in resistance and eventual breakdown owing to the onset of electromigration. Another development concerns the study of the topological insulator states and strong electron-electron correlations in quantum materials, which traditionally were studied separately. The situation changed when some quasi-1D vdW materials revealed topological phase transitions emerging due to the strongly correlated interactions, *e.g.* formation of CDW in a Weyl semimetal.

In light of these exciting developments, we conclude that much more is to come in the field of 1D quantum materials. These materials are excellent platforms for fundamental and applied experiments and theoretical studies. Theory suggests that many 1D vdW materials do not open an energy band-gap when their thickness approaches an individual atomic chain, thus creating the possibility of identifying materials that maintain low electrical resistivity all the way down to the atomic chain limit. Such materials can become a foundation of future interconnect technologies in nanoelectronics and neuromorphic networks. Furthermore, the true 1D materials are inherently susceptible to the Peierls instability. Correspondingly, many 1D crystals exfoliated into few-chain threads may reveal quantum CDW phases even if the effects are absent in bulk crystals. It is expected that many of such CDW phases will exist at RT and above, over expansive length scales. However, one needs to keep in mind that CDW phenomena may require atomic chain bundles with a certain minimum diameter. It is known in the context of 2D van der Waals materials that the commensurate CDW phase transition disappears when the thickness of the films decreases below ~9 nm [164,261]. Controlling the CDW phases in quasi-1D van der Waals materials with electrical gates will open possibilities for applications in information processing and quantum computing.

## Acknowledgements

A.A.B. is supported by the Vannevar Bush Faculty Fellowship from the Office of Secretary of Defense (OSD), under the office of Naval Research (ONR) contract N00014-21-1-2947. The work at UC Riverside, described in the review, was supported, in part, by the National Science Foundation (NSF) program Designing Materials to Revolutionize and Engineer our Future (DMREF) *via* a project DMR-1921958 entitled Collaborative Research: Data Driven Discovery of Synthesis Pathways and Distinguishing Electronic Phenomena of 1D van der Waals Bonded Solids; and by the Semiconductor Research Corporation (SRC) contract 2018-NM-2796 entitled One-Dimensional Single-Crystal van-der-Waals Metals: Ultimately-Downscaled Interconnects with Exceptional Current-Carrying Capacity and Reliability. We thank Zahra Barani Beiranvand (UC Riverside) and Yassamin Ghafouri (UGA) for their assistance with preparing the figures.

## References

[1]     L. Esaki, R. Tsu, IBM J. Res. Dev. 14 (1970) 61–65.

[2]     A.Y. Cho, J.R. Arthur, Prog. Solid State Chem. 10 (1975) 157–191.

[3]     H. Sakaki, in: Int. Conf. Mol. Beam Ep., Institute of Electrical and Electronics Engineers (IEEE) Inc., 2002, p. 5.

[4]     R.S. Wagner, W.C. Ellis, Appl. Phys. Lett. 4 (1964) 89–90.

[5]     R.S. Wagner, A.P. Levitt, Whisker Technology, 1970.

[6]     D. Leonard, K. Pond, P.M. Petroff, Phys. Rev. B 50 (1994) 11687.

[7]     D.J. Eaglesham, M. Cerullo, Phys. Rev. Lett. 64 (1990) 1943.

[8]     A. Ekimov, A. Onushchenko, JETP Lett 34 (1981) 345–349.

[9]     L.E. Brus, J. Chem. Phys. 79 (1998) 5566.

[10]    A.I. Ekimov, A.L. Efros, A.A. Onushchenko, Solid State Commun. 88 (1993) 947–950.

[11]    C.B. Murray, D.J. Norris, M.G. Bawendi, J. Am. Chem. Soc. 115 (2002) 8706–8715.

[12]    H.W. Kroto, J.R. Heath, S.C. O'Brien, R.F. Curl, R.E. Smalley, Nature 318 (1985) 162–163.

[13]    S. Iijima, T. Ichihashi, Nature 363 (1993) 603–605.

[14]    D.S. Bethune, C.H. Klang, M.S. De Vries, G. Gorman, R. Savoy, J. Vazquez, R. Beyers, Nature 363 (1993) 605–607.

[15]    V. Lukyanovich, Sov. J. Phys. Chem. 26 (1952) 88–95.

[16]    M. Monthioux, V.L. Kuznetsov, Carbon 44 (2006) 1621–1623.

[17]    K.S. Novoselov, A.K. Geim, S. V Morozov, D. Jiang, Y. Zhang, S. V Dubonos, I. V Grigorieva, A.A. Firsov, Science 306 (2004) 666–9.

[18]    Y. Zhang, Y.-W. Tan, H.L. Stormer, P. Kim, Nature 438 (2005) 201–204.

[19]    A.K. Geim, K.S. Novoselov, Nat. Mater. 6 (2007) 183–191.

[20]   S. Ghosh, W. Bao, D.L. Nika, S. Subrina, E.P. Pokatilov, C.N. Lau, A.A. Balandin, Nat. Mater. 9 (2010) 555–558.

[21]   A.A. Balandin, Nat. Mater. 10 (2011) 569–581.

[22]   F. Kargar, E.A. Coleman, S. Ghosh, J. Lee, M.J. Gomez, Y. Liu, A.S. Magana, Z. Barani, A. Mohammadzadeh, B. Debnath, R.B. Wilson, R.K. Lake, A.A. Balandin, ACS Nano 14 (2020) 2424–2435.

[23]   A.K. Geim, I. V. Grigorieva, Nature 499 (2013) 419–425.

[24]   P. Ajayan, P. Kim, K. Banerjee, Phys. Today 69 (2016) 38–44.

[25]   K.S. Novoselov, A. Mishchenko, A. Carvalho, A.H. Castro Neto, Science 353 (2016) aac9439.

[26]   M.A. Stolyarov, G. Liu, M.A. Bloodgood, E. Aytan, C. Jiang, R. Samnakay, T.T. Salguero, D.L. Nika, S.L. Rumyantsev, M.S. Shur, K.N. Bozhilov, A.A. Balandin, Nanoscale 8 (2016) 15774–15782.

[27]   G. Liu, S. Rumyantsev, M.A. Bloodgood, T.T. Salguero, M. Shur, A.A. Balandin, Nano Lett. 17 (2017) 377–383.

[28]   A. Geremew, M.A. Bloodgood, E. Aytan, B.W.K. Woo, S.R. Corber, G. Liu, K. Bozhilov, T.T. Salguero, S. Rumyantsev, M.P. Rao, A.A. Balandin, IEEE Electron Device Lett. 39 (2018) 735–738.

[29]   A.K. Geremew, S. Rumyantsev, M.A. Bloodgood, T.T. Salguero, A.A. Balandin, Nanoscale 10 (2018) 19749–19756.

[30]   M.A. Bloodgood, P. Wei, E. Aytan, K.N. Bozhilov, A.A. Balandin, T.T. Salguero, APL Mater. 6 (2018) 026602.

[31]   T.A. Empante, A. Martinez, M. Wurch, Y. Zhu, A.K. Geremew, K. Yamaguchi, M. Isarraraz, S. Rumyantsev, E.J. Reed, A.A. Balandin, L. Bartels, Nano Lett. 19 (2019) 4355–4361.

[32]   S. Baraghani, J. Abourahma, Z. Barani, A. Mohammadzadeh, S. Sudhindra, A. Lipatov, A. Sinitskii, F. Kargar, A.A. Balandin, ACS Appl. Mater. Interfaces 13 (2021) 47033–

47042.

[33]  B. Huang, G. Clark, E. Navarro-Moratalla, D.R. Klein, R. Cheng, K.L. Seyler, D. Zhong, E. Schmidgall, M.A. McGuire, D.H. Cobden, W. Yao, D. Xiao, P. Jarillo-Herrero, X. Xu, Nature 546 (2017) 270–273.

[34]  S. Niu, G. Joe, H. Zhao, Y. Zhou, T. Orvis, H. Huyan, J. Salman, K. Mahalingam, B. Urwin, J. Wu, Y. Liu, T.E. Tiwald, S.B. Cronin, B.M. Howe, M. Mecklenburg, R. Haiges, D.J. Singh, H. Wang, M.A. Kats, J. Ravichandran, Nat. Photonics 12 (2018) 392–396.

[35]  L. Yang, Y. Tao, Y. Zhu, M. Akter, K. Wang, Z. Pan, Y. Zhao, Q. Zhang, Y.Q. Xu, R. Chen, T.T. Xu, Y. Chen, Z. Mao, D. Li, Nat. Nanotechnol. 16 (2021) 764–768.

[36]  K. Xu, L. Qin, J.R. Heath, Nat. Nanotechnol. 4 (2009) 368–372.

[37]  D. Laroche, G. Gervais, M.P. Lilly, J.L. Reno, Nat. Nanotechnol. 6 (2011) 793–797.

[38]  T. Christoff-Tempesta, Y. Cho, D.Y. Kim, M. Geri, G. Lamour, A.J. Lew, X. Zuo, W.R. Lindemann, J.H. Ortony, Nat. Nanotechnol. 16 (2021) 447–454.

[39]  J. Lahiri, Y. Lin, P. Bozkurt, I.I. Oleynik, M. Batzill, Nat. Nanotechnol. 5 (2010) 326–329.

[40]  L.A. Jauregui, M.T. Pettes, L.P. Rokhinson, L. Shi, Y.P. Chen, Nat. Nanotechnol. 11 (2016) 345–351.

[41]  K. Wu, E. Torun, H. Sahin, B. Chen, X. Fan, A. Pant, D.P. Wright, T. Aoki, F.M. Peeters, E. Soignard, S. Tongay, Nat. Commun. 7 (2016) 12952.

[42]  Z. Barani, F. Kargar, Y. Ghafouri, S. Baraghani, S. Sudhindra, A. Mohammadzadeh, T.T. Salguero, A.A. Balandin, ACS Appl. Mater. Interfaces 13 (2021) 21527–21533.

[43]  Y. Qi, W. Shi, P. Werner, P.G. Naumov, W. Schnelle, L. Wang, K.G. Rana, S. Parkin, S.A. Medvedev, B. Yan, C. Felser, Npj Quantum Mater. 3 (2018) 1–6.

[44]  J.S. Rhyee, K.H. Lee, S.M. Lee, E. Cho, S. Il Kim, E. Lee, Y.S. Kwon, J.H. Shim, G. Kotliar, Nature 459 (2009) 965–968.

[45]  Z. Barani, F. Kargar, Y. Ghafouri, S. Ghosh, K. Godziszewski, S. Baraghani, Y.

Yashchyshyn, G. Cywiński, S. Rumyantsev, T.T. Salguero, A.A. Balandin, Adv. Mater. 33 (2021) 2007286.

[46] C. Lin, M. Ochi, R. Noguchi, K. Kuroda, M. Sakoda, A. Nomura, M. Tsubota, P. Zhang, C. Bareille, K. Kurokawa, Y. Arai, K. Kawaguchi, H. Tanaka, K. Yaji, A. Harasawa, M. Hashimoto, D. Lu, S. Shin, R. Arita, S. Tanda, T. Kondo, Nat. Mater. 20 (2021) 1093–1099.

[47] R. Noguchi, T. Takahashi, K. Kuroda, M. Ochi, T. Shirasawa, M. Sakano, C. Bareille, M. Nakayama, M.D. Watson, K. Yaji, A. Harasawa, H. Iwasawa, P. Dudin, T.K. Kim, M. Hoesch, V. Kandyba, A. Giampietri, A. Barinov, S. Shin, R. Arita, T. Sasagawa, T. Kondo, Nature 566 (2019) 518–522.

[48] R. Xiang, T. Inoue, Y. Zheng, A. Kumamoto, Y. Qian, Y. Sato, M. Liu, D. Tang, D. Gokhale, J. Guo, K. Hisama, S. Yotsumoto, T. Ogamoto, H. Arai, Y. Kobayashi, H. Zhang, B. Hou, A. Anisimov, M. Maruyama, Y. Miyata, S. Okada, S. Chiashi, Y. Li, J. Kong, E.I. Kauppinen, Y. Ikuhara, K. Suenaga, S. Maruyama, Science 367 (2020) 537–542.

[49] M. Endres, H. Bernien, A. Keesling, H. Levine, E.R. Anschuetz, A. Krajenbrink, C. Senko, V. Vuletic, M. Greiner, M.D. Lukin, Science 354 (2016) 1024–1027.

[50] A. Ambrosetti, N. Ferri, R.A. DiStasio, A. Tkatchenko, Science 351 (2016) 1171–1176.

[51] J. Lu, X. Xu, M. Greenblatt, R. Jin, P. Tinnemans, S. Licciardello, M.R. Van Delft, J. Buhot, P. Chudzinski, N.E. Hussey, Sci. Adv. 5 (2019) eaar8027.

[52] Y. Hu, F. Florio, Z. Chen, W. Adam Phelan, M.A. Siegler, Z. Zhou, Y. Guo, R. Hawks, J. Jiang, J. Feng, L. Zhang, B. Wang, Y. Wang, D. Gall, E.F. Palermo, Z. Lu, X. Sun, T.M. Lu, H. Zhou, Y. Ren, E. Wertz, R. Sundararaman, J. Shi, Sci. Adv. 6 (2020) eaay4213.

[53] J.O. Island, A.J. Molina-Mendoza, M. Barawi, R. Biele, E. Flores, J.M. Clamagirand, J.R. Ares, C. Sánchez, H.S.J. Van Der Zant, R. D'Agosta, I.J. Ferrer, A. Castellanos-Gomez, 2D Mater. 4 (2017) 022003.

[54] A. Patra, C.S. Rout, RSC Adv. 10 (2020) 36413–36438.

[55]    V.Y. Pokrovskii, S.G. Zybtsev, M. V Nikitin, I.G. Gorlova, V.F. Nasretdinova, S. V Zaitsev-Zotov, Physics-Uspekhi 56 (2013) 29–48.

[56]    M.N. Kozlova, A.N. Enyashin, V.E. Fedorov, J. Struct. Chem. 2016 578 57 (2017) 1505–1512.

[57]    H. Bergeron, D. Lebedev, M.C. Hersam, Chem. Rev. 121 (2021) 2713–2775.

[58]    S. Oh, S. Chae, B.J. Kim, A.J. Siddiqa, K.H. Choi, W.S. Jang, K.H. Lee, H.Y. Kim, D.K. Lee, Y.M. Kim, H.K. Yu, J.Y. Choi, Phys. Status Solidi - Rapid Res. Lett. 12 (2018) 1800451.

[59]    J. Gopalakrishnan, K.S. Nanjundaswamy, Bull. Mater. Sci. 5 (1983) 287–306.

[60]    H. Fujishita, M. Sato, S. Sato, S. Hoshino, J. Phys. C Solid State Phys. 18 (1985) 1105.

[61]    S.M. Shapiro, M. Sato, S. Hoshino, J. Phys. C Solid State Phys. 19 (1986) 3049.

[62]    P. Gressier, A. Meerschaut, L. Guemas, J. Rouxel, P. Monceau, J. Solid State Chem. 51 (1984) 141–151.

[63]    S. Liu, S.-M. Nie, J. Phys. C Solid State Phys. 12 (1979) 277.

[64]    S.K. Srivastava, B.N. Avasthi, J. Mater. Sci. 27 (1992) 3693–3705.

[65]    T. Debnath, B. Debnath, R.K. Lake, Phys. Rev. Mater. 5 (2021) 034010.

[66]    F.J. Di Salvo, C.H. Chen, R.M. Fleming, J. V Waszczak, R.G. Dunn, S.A. Sunshine, J.A. Ibers, J. Less Common Met. 116 (1986) 51–61.

[67]    S. Furuseth, L. Brattas, A. Kjekshus, Acta Chem. Scand. 27 (1973) 2367–2374.

[68]    Q.M. Liu, D. Wu, Z.A. Li, L.Y. Shi, Z.X. Wang, S.J. Zhang, T. Lin, T.C. Hu, H.F. Tian, J.Q. Li, T. Dong, N.L. Wang, Nat. Commun. 2021 121 12 (2021) 2050.

[69]    J. Qiao, F. Feng, Z. Wang, M. Shen, G. Zhang, X. Yuan, M.G. Somekh, ACS Appl. Mater. Interfaces 13 (2021) 17948–17956.

[70]    Y. Zhang, W. Yu, J. Li, J. Chen, Z. Dong, L. Xie, C. Li, X. Shi, W. Guo, S. Lin, S. Mokkapati, K. Zhang, Mater. Des. 208 (2021) 109894.

[71]  J. Niu, J. Wang, Z. He, C. Zhang, X. Li, T. Cai, X. Ma, S. Jia, D. Yu, X. Wu, Phys. Rev. B 95 (2017) 35420.

[72]  Z. Guo, H. Gu, M. Fang, B. Song, W. Wang, X. Chen, C. Zhang, H. Jiang, L. Wang, S. Liu, ACS Mater. Lett. 3 (2021) 525–534.

[73]  P. Yang, W. Wang, X. Zhang, K. Wang, L. He, W. Liu, Y. Xu, Sci. Rep. 9 (2019) 3558.

[74]  G. Cheon, K.-A.N. Duerloo, A.D. Sendek, C. Porter, Y. Chen, E.J. Reed, Nano Lett. 17 (2017) 1915–1923.

[75]  L. Fu, C. Shang, S. Zhou, Y. Guo, J. Zhao, Appl. Phys. Lett. 120 (2022) 023103.

[76]  F. Lu, J. Cui, P. Liu, M. Lin, Y. Cheng, H. Liu, W. Wang, K. Cho, W.H. Wang, Chinese Phys. B 30 (2021) 057304.

[77]  L.H. Brixner, J. Inorg. Nucl. Chem. 24 (1962) 257–263.

[78]  E. Bjerkelund, A. Kjekshus, ZAAC - J. Inorg. Gen. Chem. 328 (1964) 235–242.

[79]  P. Monceau, ed., Electronic Properties of Inorganic Quasi-One-Dimensional Compounds, Part I — Theoretical, 1st ed., Springer, Dordrecht, 1985.

[80]  S. Kagoshima, H. Nagasawa, T. Sambongi, One-Dimensional Conductors, Springer, Berlin, Heidelberg, Berlin, Heidelberg, 1988.

[81]  J. Rouxel, ed., Crystal Chemistry and Properties of Materials with Quasi-One-Dimensional Structures, A Chemical and Physical Synthetic Approach, 1st ed., Springer, Dordrecht, 1986.

[82]  D. Baeriswyl, L. Degiorgi, eds., Strong Interactions in Low Dimensions, 1st ed., Springer, Dordrecht, 2004.

[83]  T. Sambongi, M. Yamamoto, K. Tsutsumi, Y. Shiozaki, K. Yamaya, Y. Abe, J. Phys. Soc. Japan 42 (1977) 1421–1422.

[84]  P. Haen, F. Lapierre, P. Monceau, M. Núñez Regueiro, J. Richard, Solid State Commun. 26 (1978) 725–730.

[85]  M. Yamamoto, J. Phys. Soc. Japan 45 (1978) 431–438.

[86]   K. Yamaya, S. Takayanagi, S. Tanda, Phys. Rev. B - Condens. Matter Mater. Phys. 85 (2012) 184513.

[87]   G.A. Toombs, Phys. Rep. 40 (1978) 181–240.

[88]   G. Grüner, Rev. Mod. Phys. 60 (1988) 1129–1181.

[89]   G. Grüner, Density Waves in Solids, Addison-Wesley Pub. Co., Advanced Book Program, 1994.

[90]   P. Monceau, Adv. Phys. 61 (2012) 325–581.

[91]   S. Barišiĉ, A. Bjeliš, J.R. Cooper, B.A. Leontić, eds., in: Lect. Notes Phys., Springer, Berlin, Heidelberg, 1979.

[92]   S. Barišiĉ, A. Bjeliš, J.R. Cooper, B. Leontić, eds., in: Lect. Notes Phys., Springer, Berlin, Heidelberg, 1979.

[93]   M.C. Böhm, One-Dimensional Organometallic Materials, An Analysis of Electronic Structure Effects, Springer, Berlin, Heidelberg, 1987.

[94]   M.E. Itkis, F. Ya Nad', P. Monceau, J. Phys. Condens. Matter 2 (1990) 8327–8335.

[95]   S.G. Zybtsev, V.Y. Pokrovskii, V.F. Nasretdinova, S. V Zaitsev-Zotov, V. V. Pavlovskiy, A.B. Odobesco, W.W. Pai, M.W. Chu, Y.G. Lin, E. Zupanič, H.J.P. Van Midden, S. Šturm, E. Tchernychova, A. Prodan, J.C. Bennett, I.R. Mukhamedshin, O. V Chernysheva, A.P. Menushenkov, V.B. Loginov, B.A. Loginov, A.N. Titov, M. Abdel-Hafiez, Phys. Rev. B 95 (2017) 35110.

[96]   R.M. Fleming, D.E. Moncton, D.B. McWhan, Phys. Rev. B 18 (1978) 5560–5563.

[97]   A. Zettl, G. Grüner, Phys. Rev. B 26 (1982) 2298–2301.

[98]   R.P. Hall, A. Zettl, Phys. Rev. B 38 (1988) 13019–13027.

[99]   A. V. Frolov, A.P. Orlov, F. Gay, A.A. Sinchenko, P. Monceau, Appl. Phys. Lett. 118 (2021) 213102.

[100]  J. Bardeen, E. Ben-Jacob, A. Zettl, G. Grüner, Phys. Rev. Lett. 49 (1982) 493–496.

[101]  A. Zettl, C.M. Jackson, G. Grüner, Phys. Rev. B 26 (1982) 5773–5785.

[102]  R.E. Thorne, W.G. Lyons, J.W. Lyding, J.R. Tucker, J. Bardeen, Phys. Rev. B 35 (1987) 6360–6372.

[103]  H.S.J. van der Zant, E. Slot, S. V Zaitsev-Zotov, S.N. Artemenko, Phys. Rev. Lett. 87 (2001) 126401.

[104]  S.G. Zybtsev, V.Y. Pokrovskii, V.F. Nasretdinova, S. V Zaitsev-Zotov, Appl. Phys. Lett. 94 (2009) 152112.

[105]  S.G. Zybtsev, S.A. Nikonov, V.Y. Pokrovskii, V. V Pavlovskiy, D. Starešinić, Phys. Rev. B 101 (2020) 115425.

[106]  S.A. Nikonov, S.G. Zybtsev, A.A. Maizlakh, V.Y. Pokrovskii, Appl. Phys. Lett. 118 (2021) 213106.

[107]  S. V Zaitsev-Zotov, Physics-Uspekhi 47 (2004) 533–554.

[108]  J. McCarten, D.A. DiCarlo, M.P. Maher, T.L. Adelman, R.E. Thorne, Phys. Rev. B 46 (1992) 4456–4482.

[109]  H.S.J. Van Der Zant, A. Kalwij, O.C. Mantel, N. Markovic, Y.I. Latyshev, B. Pannetier, P. Monceau, J. Phys. IV  JP 9 (1999) Pr10-157.

[110]  H.J.S. van der Zant, N. Markovic, E. Slot, Physics-Uspekhi 44 (2001) 61–65.

[111]  E. Slot, H.S.J. van der Zant, K. O'neill, R.E. Thorne, Phys. Rev. B - Condens. Matter Mater. Phys. 69 (2004) 073105.

[112]  K. Inagaki, T. Toshima, S. Tanda, K. Yamaya, S. Uji, Appl. Phys. Lett. 86 (2005) 1–3.

[113]  Y.S. Hor, Z.L. Xiao, U. Welp, Y. Ito, J.F. Mitchell, R.E. Cook, W.K. Kwok, G.W. Crabtreet, Nano Lett. 5 (2005) 397–401.

[114]  S.G. Zybtsev, V.Y. Pokrovskii, Phys. Rev. B - Condens. Matter Mater. Phys. 84 (2011) 85139.

[115]  S. Bhattacharya, A.N. Bloch, J.P. Stokes, Frequency Modulator and Demodulator Using Material Having Sliding Charge Density Waves, U.S. Patent No. 4,580,110 (1985).

[116] T.L. Adelman, S. V Zaitsev-Zotov, R.E. Thorne, Phys. Rev. Lett. 74 (1995) 5264–5267.

[117] G. Blumberg, P.B. Littlewood, Electronic Devices Based on Density Wave Dielectrics, U.S. Patent No. 6,735,073 (2004), n.d.

[118] J. Demsar, J. Demsar, L. Forró, H. Berger, D. Mihailovic, Phys. Rev. B - Condens. Matter Mater. Phys. 66 (2002) 411011–411014.

[119] C. Felser, E.W. Finckh, H. Kleinke, F. Rocker, W. Tremel, J. Mater. Chem. 8 (1998) 1787–1798.

[120] T. Yokoya, T. Kiss, A. Chainani, S. Shin, K. Yamaya, Phys. Rev. B - Condens. Matter Mater. Phys. 71 (2005) 140504(R).

[121] M. Hoesch, A. Bosak, D. Chernyshov, H. Berger, M. Krisch, Phys. Rev. Lett. 102 (2009) 086402.

[122] M. Hoesch, X. Cui, K. Shimada, C. Battaglia, S.I. Fujimori, H. Berger, Phys. Rev. B - Condens. Matter Mater. Phys. 80 (2009) 075423.

[123] Y. Hu, F. Zheng, X. Ren, J. Feng, Y. Li, Phys. Rev. B - Condens. Matter Mater. Phys. 91 (2015) 144502.

[124] S.L. Gleason, Y. Gim, T. Byrum, A. Kogar, P. Abbamonte, E. Fradkin, G.J. Macdougall, D.J. Van Harlingen, X. Zhu, C. Petrovic, S.L. Cooper, Phys. Rev. B - Condens. Matter Mater. Phys. 91 (2015) 155124.

[125] A.M. Ganose, L. Gannon, F. Fabrizi, H. Nowell, S.A. Barnett, H. Lei, X. Zhu, C. Petrovic, D.O. Scanlon, M. Hoesch, Phys. Rev. B 97 (2018) 155103.

[126] M. Hoesch, L. Gannon, K. Shimada, B.J. Parrett, M.D. Watson, T.K. Kim, X. Zhu, C. Petrovic, Phys. Rev. Lett. 122 (2019) 017601.

[127] L. Yue, S. Xue, J. Li, W. Hu, A. Barbour, F. Zheng, L. Wang, J. Feng, S.B. Wilkins, C. Mazzoli, R. Comin, Y. Li, Nat. Commun. 11 (2020) 98.

[128] X. Liu, J. Liu, L.Y. Antipina, J. Hu, C. Yue, A.M. Sanchez, P.B. Sorokin, Z. Mao, J. Wei, Nano Lett. 16 (2016) 6188–6195.

[129]  J. Dai, M. Li, X.C. Zeng, Wiley Interdiscip. Rev. Comput. Mol. Sci. 6 (2016) 211–222.

[130]  M.D. Randle, A. Lipatov, I. Mansaray, J.E. Han, A. Sinitskii, J.P. Bird, Appl. Phys. Lett. 118 (2021) 210502.

[131]  J.O. Island, R. Biele, M. Barawi, J.M. Clamagirand, J.R. Ares, C. Sánchez, H.S.J. Van Der Zant, I.J. Ferrer, R. D'Agosta, A. Castellanos-Gomez, Sci. Rep. 6 (2016) 22214.

[132]  J.O. Island, M. Barawi, R. Biele, A. Almazán, J.M. Clamagirand, J.R. Ares, C. Sánchez, H.S.J. Van Der Zant, J. V. Álvarez, R. D'Agosta, I.J. Ferrer, A. Castellanos-Gomez, Adv. Mater. 27 (2015) 2595–2601.

[133]  A. Lipatov, P.M. Wilson, M. Shekhirev, J.D. Teeter, R. Netusil, A. Sinitskii, Nanoscale 7 (2015) 12291–12296.

[134]  S.J. Gilbert, A. Lipatov, A.J. Yost, M.J. Loes, A. Sinitskii, P.A. Dowben, Appl. Phys. Lett. 114 (2019) 101604.

[135]  M. Randle, A. Lipatov, A. Kumar, C.P. Kwan, J. Nathawat, B. Barut, S. Yin, K. He, N. Arabchigavkani, R. Dixit, T. Komesu, J. Avila, M.C. Asensio, P.A. Dowben, A. Sinitskii, U. Singisetti, J.P. Bird, ACS Nano 13 (2019) 803–811.

[136]  F. Iyikanat, R.T. Senger, F.M. Peeters, H. Sahin, ChemPhysChem 17 (2016) 3985–3991.

[137]  N. Papadopoulos, E. Flores, K. Watanabe, T. Taniguchi, J.R. Ares, C. Sanchez, I.J. Ferrer, A. Castellanos-Gomez, G.A. Steele, H.S.J. Van Der Zant, 2D Mater. 7 (2019) 015009.

[138]  S.S. Sylvia, R.K. Lake, ArXiv:1810.07734 [Cond-Mat.Mtrl-Sci] (2018).

[139]  D.A. Bandurin, A. V. Tyurnina, G.L. Yu, A. Mishchenko, V. Zólyomi, S. V. Morozov, R.K. Kumar, R. V. Gorbachev, Z.R. Kudrynskyi, S. Pezzini, Z.D. Kovalyuk, U. Zeitler, K.S. Novoselov, A. Patanè, L. Eaves, I. V. Grigorieva, V.I. Fal'Ko, A.K. Geim, Y. Cao, Nat. Nanotechnol. 12 (2017) 223–227.

[140]  H. Yi, T. Komesu, S. Gilbert, G. Hao, A.J. Yost, A. Lipatov, A. Sinitskii, J. Avila, C. Chen, M.C. Asensio, P.A. Dowben, Appl. Phys. Lett. 112 (2018) 052102.

[141]  J.A. Silva-Guillén, E. Canadell, P. Ordejón, F. Guinea, R. Roldán, 2D Mater. 4 (2017) 025085.

[142]  S. Takahashi, T. Sambongi, J.W. Brill, W. Roark, Solid State Commun. 49 (1984) 1031–1034.

[143]  A. Sahu, S.N. Steinmann, P. Raybaud, Cryst. Growth Des. 20 (2020) 7750–7760.

[144]  W.W. Fuller, P.M. Chaikin, N.P. Ong, Solid State Commun. 30 (1979) 689–692.

[145]  K. Nishida, T. Sambongi, M. Ido, J. Phys. Soc. Japan 48 (1980) 331–332.

[146]  X.Z. Chen, J.Q. Shen, Y. Xu, J. Zhou, Z.A. Xu, Phys. B Condens. Matter 352 (2004) 280–284.

[147]  D.S. Muratov, V.O. Vanyushin, N.S. Vorobeva, P. Jukova, A. Lipatov, E.A. Kolesnikov, D. Karpenkov, D. V. Kuznetsov, A. Sinitskii, J. Alloys Compd. 815 (2020) 152316.

[148]  P.R.N. Misse, D. Berthebaud, O.I. Lebedev, A. Maignan, E. Guilmeau, Materials 8 (2015) 2514–2522.

[149]  E. Flores, J.R. Ares, C. Sánchez, I.J. Ferrer, Catal. Today 321–322 (2019) 107–112.

[150]  K. Wu, M. Blei, B. Chen, L. Liu, H. Cai, C. Brayfield, D. Wright, H. Zhuang, S. Tongay, K. Wu, M. Blei, B. Chen, L. Liu, H. Cai, C. Brayfield, H. Zhuang, S. Tongay, D. Wright, Adv. Mater. 32 (2020) 2000018.

[151]  M.S. Whittingham, J. Electroanal. Chem. Interfacial Electrochem. 118 (1981) 229–239.

[152]  F. Iyikanat, H. Sahin, R.T. Senger, F.M. Peeters, J. Phys. Chem. C 119 (2015) 10709–10715.

[153]  Z. Tian, C. Han, Y. Zhao, W. Dai, X. Lian, Y. Wang, Y. Zheng, Y. Shi, X. Pan, Z. Huang, H. Li, W. Chen, Nat. Commun. 12 (2021) 2039.

[154]  Y. Wei, Z. Zhou, R. Long, J. Phys. Chem. Lett. 8 (2017) 4522–4529.

[155]  Z. Tian, X. Guo, D. Wang, D. Sun, S. Zhang, K. Bu, W. Zhao, F. Huang, Z. Tian, X. Guo, D. Wang, D. Sun, S. Zhang, K. Bu, W. Zhao, F. Huang, Adv. Funct. Mater. 30 (2020) 2001286.

[156]  S. Conejeros, B. Guster, P. Alemany, J.P. Pouget, E. Canadell, Chem. Mater. 33 (2021) 5449–5463.

[157]   N. Tripathi, V. Pavelyev, P. Sharma, S. Kumar, A. Rymzhina, P. Mishra, Mater. Sci. Semicond. Process. 127 (2021) 105699.

[158]   J.Á. Silva-Guillén, E. Canadell, 2D Mater. 7 (2020) 025038.

[159]   D.X. Qu, Y.S. Hor, J. Xiong, R.J. Cava, N.P. Ong, Science 329 (2010) 821–824.

[160]   Y.L. Chen, J.G. Analytis, J.H. Chu, Z.K. Liu, S.K. Mo, X.L. Qi, H.J. Zhang, P.H. Lu, X. Dai, Z. Fang, S.C. Zhang, I.R. Fisher, Z. Hussain, Z.X. Shen, Science 325 (2009) 178– 181.

[161]   K.M.F. Shahil, M.Z. Hossain, D. Teweldebrhan, A.A. Balandin, Appl. Phys. Lett. 96 (2010) 153103.

[162]   K.M.F. Shahil, M.Z. Hossain, V. Goyal, A.A. Balandin, J. Appl. Phys. 111 (2012) 054305.

[163]   Y. Yu, F. Yang, X.F. Lu, Y.J. Yan, Y.H. Cho, L. Ma, X. Niu, S. Kim, Y.W. Son, D. Feng, S. Li, S.W. Cheong, X.H. Chen, Y. Zhang, Nat. Nanotechnol. 10 (2015) 270–276.

[164]   M. Yoshida, Y. Zhang, J. Ye, R. Suzuki, Y. Imai, S. Kimura, A. Fujiwara, Y. Iwasa, Sci. Rep. 4 (2014) 7302.

[165]   G. Liu, B. Debnath, T.R. Pope, T.T. Salguero, R.K. Lake, A.A. Balandin, Nat Nano 11 (2016) 845–850.

[166]   Y. Li, Y. Rao, K.F. Mak, Y. You, S. Wang, C.R. Dean, T.F. Heinz, Nano Lett. 13 (2013) 3329–3333.

[167]   L.M. Malard, T. V Alencar, A.P.M. Barboza, K.F. Mak, A.M. De Paula, Phys. Rev. B - Condens. Matter Mater. Phys. 87 (2013) 201401.

[168]   X. Yin, Z. Ye, D.A. Chenet, Y. Ye, K. O'Brien, J.C. Hone, X. Zhang, Science 344 (2014) 488–490.

[169]   G. Cheon, E.D. Cubuk, E.R. Antoniuk, L. Blumberg, J.E. Goldberger, E.J. Reed, J. Phys. Chem. Lett. 9 (2018) 6967–6972.

[170]   T. Pham, S. Oh, P. Stetz, S. Onishi, C. Kisielowski, M.L. Cohen, A. Zettl, Science 361

(2018) 263–266.

[171] S. Stonemeyer, J.D. Cain, S. Oh, A. Azizi, M. Elasha, M. Thiel, C. Song, P. Ercius, M.L. Cohen, A. Zettl, J. Am. Chem. Soc. 143 (2021) 4563–4568.

[172] Y. Jin, X. Li, J. Yang, Phys. Chem. Chem. Phys. 17 (2015) 18665–18669.

[173] W. Wang, S. Dai, X. Li, J. Yang, D.J. Srolovitz, Q. Zheng, Nat. Commun. 6 (2015) 7853.

[174] A. Lipatov, M.J. Loes, H. Lu, J. Dai, P. Patoka, N.S. Vorobeva, D.S. Muratov, G. Ulrich, B. Kästner, A. Hoehl, G. Ulm, X.C. Zeng, E. Rühl, A. Gruverman, P.A. Dowben, A. Sinitskii, ACS Nano 12 (2018) 12713–12720.

[175] A. Nomura, K. Yamaya, S. Takayanagi, K. Ichimura, S. Tanda, EPL (Europhysics Lett. 124 (2019) 67001.

[176] S.G. Zybtsev, V.Y. Pokrovskii, S. V Zaitsev-Zotov, Nat. Commun. 1 (2010) 1087.

[177] G. Grüner, A. Zawadowski, P.M. Chaikin, Phys. Rev. Lett. 46 (1981) 511–515.

[178] G. Grüner, A. Zettl, W.G. Clark, A.H. Thompson, Phys. Rev. B 23 (1981) 6813–6815.

[179] D. Borodin, S. V Zaitsev-Zotov, F. Nad, Zh. Eksp. Teor. Fiz 93 (1987) 1394–1409.

[180] M. Abdel-Hafiez, R. Thiyagarajan, A. Majumdar, R. Ahuja, W. Luo, A.N. Vasiliev, A.A. Maarouf, S.G. Zybtsev, V.Y. Pokrovskii, S. V Zaitsev-Zotov, V. V Pavlovskiy, W.W. Pai, W. Yang, L. V Kulik, Phys. Rev. B 99 (2019) 235126.

[181] I.A. Cohn, S.G. Zybtsev, A.P. Orlov, S. V Zaitsev-Zotov, JETP Lett. 112 (2020) 88–94.

[182] A.A. Balandin, S. V Zaitsev-Zotov, G. Grüner, Appl. Phys. Lett. 119 (2021) 170401.

[183] E. Slot, M.A. Holst, H.S.J. Van Der Zant, S. V Zaitsev-Zotov, Phys. Rev. Lett. 93 (2004) 176602.

[184] A. V. Zavalko, S. V Zaitsev-Zotov, J. Phys. IV 131 (2005) 359–360.

[185] S.G. Zybtsev, V.Y. Pokrovskii, S. V Zaitsev-Zotov, Phys. B Condens. Matter 407 (2012) 1810–1812.

[186] V.E. Minakova, V.F. Nasretdinova, S. V Zaitsev-Zotov, Phys. B Condens. Matter 460

(2015) 185–190.

[187] M.Z. Hasan, C.L. Kane, Rev. Mod. Phys. 82 (2010) 3045–3067.

[188] G. Autès, A. Isaeva, L. Moreschini, J.C. Johannsen, A. Pisoni, R. Mori, W. Zhang, T.G. Filatova, A.N. Kuznetsov, L. Forró, W. Van Den Broek, Y. Kim, K.S. Kim, A. Lanzara, J.D. Denlinger, E. Rotenberg, A. Bostwick, M. Grioni, O. V. Yazyev, Nat. Mater. 15 (2016) 154–158.

[189] A. Lau, J. Van Den Brink, C. Ortix, Phys. Rev. B 94 (2016) 165164.

[190] J. Gooth, B. Bradlyn, S. Honnali, C. Schindler, N. Kumar, J. Noky, Y. Qi, C. Shekhar, Y. Sun, Z. Wang, B.A. Bernevig, C. Felser, Nature 575 (2019) 315–319.

[191] W. Shi, B.J. Wieder, H.L. Meyerheim, Y. Sun, Y. Zhang, Y. Li, L. Shen, Y. Qi, L. Yang, J. Jena, P. Werner, K. Koepernik, S. Parkin, Y. Chen, C. Felser, B.A. Bernevig, Z. Wang, Nat. Phys. 17 (2021) 381–387.

[192] T. Konstantinova, L. Wu, W.G. Yin, J. Tao, G.D. Gu, X.J. Wang, J. Yang, I.A. Zaliznyak, Y. Zhu, Npj Quantum Mater. 5 (2020) 80.

[193] J. Lienig, in: Proc. Int. Symp. Phys. Des., 2013, pp. 33–40.

[194] J.P. Gambino, T.C. Lee, F. Chen, T.D. Sullivan, in: Proc. Int. Symp. Phys. Fail. Anal. Integr. Circuits, IPFA, APS, 2009, pp. 677–684.

[195] A.A. Balandin, D.L. Nika, Mater. Today 15 (2012) 266–275.

[196] D.L. Nika, A.A. Balandin, Reports Prog. Phys. 80 (2017) 036502.

[197] A. Dhar, Phys. Rev. Lett. 86 (2001) 5882–5885.

[198] A. Casher, J.L. Lebowitz, J. Math. Phys. 12 (1971) 1701–1711.

[199] O. Narayan, S. Ramaswamy, Phys. Rev. Lett. 89 (2002) 200601.

[200] C.W. Chang, D. Okawa, H. Garcia, A. Majumdar, A. Zettl, Phys. Rev. Lett. 101 (2008) 075903.

[201] G. Basile, C. Bernardin, S. Olla, Phys. Rev. Lett. 96 (2006) 204303.

[202] E. Fermi, P. Pasta, S. Ulam, M. Tsingou, Studies of Nonlinear Problems, Los Alamos, NM (United States), 1955.

[203] G. Fugallo, A. Cepellotti, L. Paulatto, M. Lazzeri, N. Marzari, F. Mauri, Nano Lett. 14 (2014) 6109–6114.

[204] Z. Aksamija, I. Knezevic, Appl. Phys. Lett. 98 (2011) 141919.

[205] A.A. Balandin, ACS Nano 14 (2020) 5170–5178.

[206] S. Lee, D. Broido, K. Esfarjani, G. Chen, Nat. Commun. 6 (2015) 6290.

[207] A. Cepellotti, G. Fugallo, L. Paulatto, M. Lazzeri, F. Mauri, N. Marzari, Nat. Commun. 6 (2015) 6400.

[208] S. Huberman, R.A. Duncan, K. Chen, B. Song, V. Chiloyan, Z. Ding, A.A. Maznev, G. Chen, K.A. Nelson, Science 364 (2019) 375–379.

[209] Y. Machida, A. Subedi, K. Akiba, A. Miyake, M. Tokunaga, Y. Akahama, K. Izawa, K. Behnia, Sci. Adv. 4 (2018) eaat3374.

[210] L. Landau, Phys. Rev. 60 (1941) 356–358.

[211] E.W. Prohofsky, J.A. Krumhansl, Phys. Rev. 133 (1964) A1403.

[212] Z. Ding, J. Zhou, B. Song, V. Chiloyan, M. Li, T.H. Liu, G. Chen, Nano Lett. 18 (2018) 638–649.

[213] Y. Machida, N. Matsumoto, T. Isono, K. Behnia, Science 367 (2020) 309–312.

[214] R. Klein, in: Nonequilibrium Phonon Dyn., Springer US, 1985, pp. 313–355.

[215] F. Kargar, B. Debnath, J.-P. Kakko, A. Saÿnätjoki, H. Lipsanen, D.L. Nika, R.K. Lake, A.A. Balandin, Nat. Commun. 7 (2016) 13400.

[216] D.L. Nika, A.I. Cocemasov, C.I. Isacova, A.A. Balandin, V.M. Fomin, O.G. Schmidt, Phys. Rev. B 85 (2012) 205439.

[217] A.I. Cocemasov, D.L. Nika, A.A. Balandin, Phys. Rev. B 88 (2013) 35428.

[218] H. Li, H. Ying, X. Chen, D.L. Nika, A.I. Cocemasov, W. Cai, A.A. Balandin, S. Chen,

Nanoscale 6 (2014) 13402–13408.

[219]  D.L. Nika, A.I. Cocemasov, A.A. Balandin, Appl. Phys. Lett. 105 (2014) 031904.

[220]  Q. Zhang, C. Liu, X. Liu, J. Liu, Z. Cui, Y. Zhang, L. Yang, Y. Zhao, T.T. Xu, Y. Chen, J. Wei, Z. Mao, D. Li, ACS Nano 12 (2018) 2634–2642.

[221]  D. Yang, W. Yao, Y. Yan, W. Qiu, L. Guo, X. Lu, C. Uher, X. Han, G. Wang, T. Yang, X. Zhou, NPG Asia Mater. 9 (2017) e387–e387.

[222]  S. Misra, C. Barreteau, J.-C. Crivello, V.M. Giordano, J.-P. Castellan, Y. Sidis, P. Levinský, J. Hejtmánek, B. Malaman, A. Dauscher, B. Lenoir, C. Candolfi, S. Pailhès, Phys. Rev. Res. 2 (2020) 043371.

[223]  M.K. Jana, K. Pal, U. V. Waghmare, K. Biswas, Angew. Chemie - Int. Ed. 55 (2016) 7792–7796.

[224]  Z. Zhou, H. Liu, D. Fan, G. Cao, C. Sheng, ACS Appl. Mater. Interfaces 10 (2018) 37031–37037.

[225]  G.J. Snyder, E.S. Toberer, Nat. Mater. 7 (2008) 105–114.

[226]  T. Sakuma, S. Nishino, M. Miyata, M. Koyano, J. Electron. Mater. 47 (2018) 3177–3183.

[227]  A. Smontara, K. Biljaković, S.N. Artemenko, Phys. Rev. B 48 (1993) 4329–4334.

[228]  A. Smontara, I. Tkalčec, A. Bilušić, M. Budimir, H. Berger, Phys. B Condens. Matter 316–317 (2002) 279–282.

[229]  H. Liu, X. Yu, K. Wu, Y. Gao, S. Tongay, A. Javey, L. Chen, J. Hong, J. Wu, Nano Lett. 20 (2020) 5221–5227.

[230]  L. Yang, Y. Tao, J. Liu, C. Liu, Q. Zhang, M. Akter, Y. Zhao, T.T. Xu, Y. Xu, Z. Mao, Y. Chen, D. Li, Nano Lett. 19 (2018) 415–421.

[231]  F. Kargar, Z. Barani, R. Salgado, B. Debnath, J.S. Lewis, E. Aytan, R.K. Lake, A.A. Balandin, ACS Appl. Mater. Interfaces 10 (2018) 37555–37565.

[232]  Z. Barani, A. Mohammadzadeh, A. Geremew, C.Y. Huang, D. Coleman, L. Mangolini, F. Kargar, A.A. Balandin, Adv. Funct. Mater. 30 (2020) 1904008.

[233] S. Sudhindra, F. Rashvand, D. Wright, Z. Barani, A.D. Drozdov, S. Baraghani, C. Backes, F. Kargar, A.A. Balandin, ACS Appl. Mater. Interfaces 13 (2021) 53073–53082.

[234] F. Kargar, Z. Barani, M. Balinskiy, A.S. Magana, J.S. Lewis, A.A. Balandin, Adv. Electron. Mater. 5 (2019) 1800558.

[235] S. Naghibi, F. Kargar, D. Wright, C.Y.T. Huang, A. Mohammadzadeh, Z. Barani, R. Salgado, A.A. Balandin, Adv. Electron. Mater. 6 (2020) 1901303.

[236] Z. Barani, F. Kargar, K. Godziszewski, A. Rehman, Y. Yashchyshyn, S. Rumyantsev, G. Cywiński, W. Knap, A.A. Balandin, ACS Appl. Mater. Interfaces 12 (2020) 28635–28644.

[237] Z. Barani, F. Kargar, A. Mohammadzadeh, S. Naghibi, C. Lo, B. Rivera, A.A. Balandin, Adv. Electron. Mater. 6 (2020) 2000520.

[238] S. Bidoki, D. McGorman, D. Lewis, M. Clark, G. Horler, R.E. Miles, AATCC Rev. 5 (2005) 11–14.

[239] U. Zschieschang, H. Klauk, J. Mater. Chem. C 7 (2019) 5522–5533.

[240] T. Carey, S. Cacovich, G. Divitini, J. Ren, A. Mansouri, J.M. Kim, C. Wang, C. Ducati, R. Sordan, F. Torrisi, Nat. Commun. 8 (2017) 1202.

[241] S. Jeong, H.C. Song, W.W. Lee, Y. Choi, S.S. Lee, B.-H. Ryu, J. Phys. Chem. C 114 (2010) 22277–22283.

[242] A. Kosmala, R. Wright, Q. Zhang, P. Kirby, Mater. Chem. Phys. 129 (2011) 1075–1080.

[243] A. Määttänen, P. Ihalainen, P. Pulkkinen, S. Wang, H. Tenhu, J. Peltonen, ACS Appl. Mater. Interfaces 4 (2012) 955–964.

[244] S. Jeong, S.H. Lee, Y. Jo, S.S. Lee, Y.-H. Seo, B.W. Ahn, G. Kim, G.-E. Jang, J.-U. Park, B.-H. Ryu, Y. Choi, J. Mater. Chem. C 1 (2013) 2704.

[245] A. Capasso, A.E. Del Rio Castillo, H. Sun, A. Ansaldo, V. Pellegrini, F. Bonaccorso, Solid State Commun. 224 (2015) 53–63.

[246] R.P. Tortorich, J.-W. Choi, Nanomater. 3 (2013) 453–468.

[247] J. Li, M.M. Naiini, S. Vaziri, M.C. Lemme, M. Östling, Adv. Funct. Mater. 24 (2014) 6524–6531.

[248] B. Cho, J. Yoon, S.K. Lim, A.R. Kim, D.-H. Kim, S.-G. Park, J.-D. Kwon, Y.-J. Lee, K.-H. Lee, B.H. Lee, H.C. Ko, M.G. Hahm, ACS Appl. Mater. Interfaces 7 (2015) 16775–16780.

[249] S.J. Rowley-Neale, C.W. Foster, G.C. Smith, D.A.C. Brownson, C.E. Banks, Sustain. Energy Fuels 1 (2017) 74–83.

[250] A.G. Kelly, T. Hallam, C. Backes, A. Harvey, A.S. Esmaeily, I. Godwin, J. Coelho, V. Nicolosi, J. Lauth, A. Kulkarni, S. Kinge, L.D.A. Siebbeles, G.S. Duesberg, J.N. Coleman, Science 356 (2017) 69–73.

[251] F. Molina-Lopez, T.Z. Gao, U. Kraft, C. Zhu, T. Öhlund, R. Pfattner, V.R. Feig, Y. Kim, S. Wang, Y. Yun, Z. Bao, Nat. Commun. 10 (2019) 2676.

[252] Z. Lin, Y. Huang, X. Duan, Nat. Electron. 2 (2019) 378–388.

[253] J.W.T. Seo, J. Zhu, V.K. Sangwan, E.B. Secor, S.G. Wallace, M.C. Hersam, ACS Appl. Mater. Interfaces 11 (2019) 5675–5681.

[254] S. Il Park, Y. Xiong, R.H. Kim, P. Elvikis, M. Meitl, D.H. Kim, J. Wu, J. Yoon, Y. Chang-Jae, Z. Liu, Y. Huang, K.C. Hwang, P. Ferreira, L. Xiuling, K. Choquette, J.A. Rogers, Science 325 (2009) 977–981.

[255] D.M. Bigg, D.E. Stutz, Polym. Compos. 4 (1983) 40–46.

[256] Y. Pan, G.J. Weng, S.A. Meguid, W.S. Bao, Z.H. Zhu, A.M.S. Hamouda, J. Appl. Phys. 110 (2011) 123715.

[257] Q. Zhao, Y. Guo, Y. Zhou, Z. Yao, Z. Ren, J. Bai, X. Xu, Nanoscale 10 (2018) 3547–3555.

[258] V.O. Özçelik, J.G. Azadani, C. Yang, S.J. Koester, T. Low, Phys. Rev. B 94 (2016) 35125.

[259] D. Wickramaratne, F. Zahid, R.K. Lake, J. Appl. Phys. 118 (2015) 075101.

[260]  D. Wickramaratne, F. Zahid, R.K. Lake, J. Chem. Phys. 140 (2014) 124710.

[261]  G. Liu, B. Debnath, T.R. Pope, T.T. Salguero, R.K. Lake, A.A. Balandin, Nat. Nanotechnol. 11 (2016) 845–850.